\documentclass{iopart}
\usepackage{iopams,graphicx}
\newcommand{\be}{\begin{eqnarray}}
\newcommand{\ee}{\end{eqnarray}}

\renewcommand{\d}{\mbox{${\rm d}$}}

\begin{document}
\title{Method of comparison equations for cosmological perturbations}
\author{R~Casadio$^1$, F~Finelli$^{2,3}$, A~Kamenshchik$^{1,4}$, 
M~Luzzi$^1$, and G~Venturi$^1$}
\address{$^1$ Dipartimento di Fisica, Universit\`a di Bologna and I.N.F.N.,
Sezione di Bologna, via Irnerio~46, 40126~Bologna, Italy}
\address{$^2$ INAF/IASF~BO, Istituto di Astrofisica Spaziale e Fisica 
Cosmica di Bologna, Istituto Nazionale di Astrofisica,
via~Gobetti~101, 40129~Bologna, Italy}
\address{$^3$ INAF/OAB, Osservatorio Astronomico di Bologna,
Istituto Nazionale di Astrofisica,
via~Ranzani~1, 40127~Bologna, Italy}
\address{$^4$ L.D.~Landau Institute for Theoretical Physics,
Russian Academy of Sciences,
Kosygin str.~2, 119334~Moscow, Russia}
\eads{\mailto{Roberto.Casadio@bo.infn.it}, 
\mailto{finelli@iasfbo.inaf.it}, 
\mailto{Alexander.Kamenshchik@bo.infn.it},
\mailto{Mattia.Luzzi@bo.infn.it},
\mailto{armitage@bo.infn.it}}
\begin{abstract}
We apply the {\em method of comparison equations} to study cosmological
perturbations during inflation, obtaining the full power spectra of
scalar and tensor perturbations to first and to second order in the
slow-roll parameters.
We compare our results with those derived by means of other methods,
in particular the Green's function method and the improved
WKB approximation, and find agreement for the slow-roll structure.
The method of comparison equations, just as the improved WKB approximation,
can however be applied to more general situations where the slow-roll
approximation fails.
\end{abstract}
\maketitle
\section{Introduction}
Anisotropies in the cosmic microwave background (CMB)
radiation and inhomogeneities in the large scale structures
of the Universe have nowadays become a fundamental tool
to study the early universe~\cite{infla1}.
Present and future data will allow us to discriminate among different
inflationary models in the near future.
For this reason, the comparison of observations with inflationary models
requires theoretical advances in the predictions of the power
spectrum of primordial perturbations beyond the lowest order
in the slow-roll parameters $\epsilon_i$'s, first obtained by Stewart and 
Lyth~\cite{SL} (their definitions will be recalled in Section~\ref{sCP}).
\par
An analytic form for the {\em full\/} inflationary power spectra to second
order in the slow-roll parameters was first obtained through the Green's
function method (GFM henceforth) in Ref.~\cite{gongstewart}.
In this way a characterization of the power spectrum to second order
in the slow-roll parameters was given which was not just derived from the
running of the spectral index (whose leading order is precisely
${\cal O}(\epsilon^2)$~\cite{KT}).
An equivalent second order characterization has been recently obtained 
by means of the improved WKB approximation of Refs.~\cite{WKB_PLB,WKB_lungo},
which extended to second order in the slow-roll parameters previous
results based on a more standard WKB approach~\cite{MS,WKB1}.
The WKB approximation has confirmed the structure of the second order 
power spectra found with the GFM, within a numerical difference in the
${\cal O}(\epsilon^2)$ coefficients of $5\,\%$ at most~\cite{WKB_PLB}.
Compared with the GFM, the WKB approximation has the additional advantage
that the slow-roll parameters do not have to be constant in
time~\cite{WKB_lungo} and can therefore be applied to a wider class
of inflationary models.
\par
The purpose of this paper is to illustrate the use of the {\em method of
comparison equations\/} (MCE in brief)~\cite{MG,dingle,berry} to predict
inflationary power spectra.
We shall see that this method yields exact results for the case of
constant slow-roll parameters (e.g.~power-law inflation) and polynomial
structures to second order in the $\epsilon_i$'s in agreement with the GFM
and WKB approximation.
The MCE however has the advantage of being more accurate to lowest
(leading) order, whereas other methods (namely, the WKB~\cite{MS}, improved
WKB~\cite{WKB1} and GFM~\cite{gongstewart}) reach a similar accuracy to
next-to-leading order.
We shall also discuss cases for which our present method appears more
flexible than the slow-roll approximation.
\par
In the next Section we briefly review the general MCE and
in Section~\ref{sCP} the theory of cosmological perturbations.
We then apply the MCE to cosmological perturbations in
Section~\ref{sMCECP} where we also analyze the error around the
``turning point'' in detail.
In Section~\ref{s_app} we analyze power-law inflation, chaotic
inflation and the arctan model;
we expand our general results to second slow-roll order and
compare with analogous results obtained with the GFM.  
We finally comment on our results in Section~\ref{sC}.
Some more technical details are given in two Appendices.
\section{Method of comparison equations}
\label{sMCE}
Let us briefly review the MCE (the name is due to Dingle~\cite{dingle}).
It was independently introduced in Refs.~\cite{MG,dingle} and
applied to wave mechanics by Berry and Mount in Ref.~\cite{berry}.
The standard WKB approximation and its improvement by Langer~\cite{langer}
are just particular cases of this method and, recently, its connection with
the Ermakov-Pinney equation was also studied~\cite{KLV}.
\par
Let us consider the second-order differential equation
\be
\left[
\frac{{\d}^2}{{\d}x^2}+\omega^2(x)
\right]
\,\chi(x)=0
\ ,
\label{exact_EQ}
\ee
where $\omega^2$ is a (not necessarily positive) ``potential''
(or ``frequency''), and suppose that we know an exact solution to a
similar second-order differential equation,
\be
\left[
\frac{{\d}^2}{{\d}\sigma^2}+\Theta^2(\sigma)
\right]
\,U(\sigma)=0
\ ,
\label{aux_EQ}
\ee
where $\Theta$ is the ``comparison function''.
One can then represent an exact solution of Eq.~(\ref{exact_EQ})
in the form
\be
\chi(x)=\left(\frac{{\d}\sigma}{{\d}x}\right)^{-1/2}\,U(\sigma)
\ ,
\label{exact_SOL}
\ee
provided the variables $x$ and $\sigma$ are related by
\be
\omega^2(x)\!=\!\left(\frac{{\d}\sigma}{{\d}x}\right)^{2}\Theta^2(\sigma)
-\left(\frac{{\d}\sigma}{{\d}x}\right)^{1/2}
\frac{{\d}^2}{{\d}x^2}\left(\frac{{\d}\sigma}{{\d}x}\right)^{-1/2}
\ .
\label{new_EQ}
\ee
Eq.~(\ref{new_EQ}) can be solved by using some iterative
scheme, in general cases~\cite{KLV,hecht} or for specific
problems~\cite{mori,pechukas}.
If we choose the comparison function sufficiently similar to
$\omega$, the second term in the right hand side (r.h.s.) of
Eq.~(\ref{new_EQ}) will be negligible with respect to the first
one, so that
\be
\omega^2(x)\simeq\left(\frac{{\d}\sigma}{{\d}x}\right)^{2}
\Theta^2(\sigma)
\ .
\label{new_EQ_appr}
\ee
On selecting a pair of values $x_0$ and $\sigma_0$ such that
$\sigma_0=\sigma(x_0)$, the function $\sigma(x)$ can be implicitly
expressed as
\be
-\xi(x)\equiv
\int_{x_0}^x\sqrt{\pm\,\omega^2(y)}\,{\d}\,y
\simeq
\int_{\sigma_0}^{\sigma}\sqrt{\pm\,\Theta^2(\rho)}\,{\d}\,\rho
\ ,
\label{new_EQ_int}
\ee
where the signs are chosen conveniently~\footnote{We recall that
$\xi(x)$ is the same quantity as used in 
Ref.~\cite{WKB1,WKB_PLB,WKB_lungo}.}.
The result in Eq.~(\ref{exact_SOL}) leads to a uniform approximation
for $\chi(x)$, valid in the whole range of the variable $x$, including
``turning points''~\footnote{With this term, borrowed from point
particle mechanics, one usually means a real zero of the ``frequency''
$\omega$.}.
The similarity between $\Theta$ and $\omega$ is clearly very
important in order to implement this method.
\section{Cosmological perturbations}
\label{sCP}
Let us begin by recalling that scalar (density) and tensor
(gravitational wave) fluctuations on a flat Robertson-Walker
background with scale factor $a$ 
\be
\d s^2=a^2\,\left(-\d\eta^2 + \d r^2 + r^2 \d\Omega^2\right)
\ ,
\ee
are given respectively by $\mu=\mu_{\rm S}\equiv a\,Q$
and $\mu=\mu_{\rm T}\equiv a\,h$, where $Q$ is the Mukhanov
variable~\cite{mukh1,mukh2} and $h$ the amplitude of the two
polarizations of gravitational waves~\cite{gris,staro}.
The functions $\mu$ must satisfy the one-dimensional
Schr\"odinger-like equation
\be
\left[\frac{{\d}^2}{{\d}\eta^2}+\Omega^2(k,\eta)\right]
\,\mu=0
\ ,
\label{osci}
\ee
together with the initial condition (corresponding to
a Bunch-Davies vacuum)
\be
\lim_{\frac{k}{a\,H}\rightarrow +\infty} \mu(k,\eta)
\simeq\frac{{\rm e}^{-i\,k\,\eta}}{\sqrt{2\,k}}
\ .
\label{init_cond_on_mu}
\ee
In the above $\eta$ is the conformal time
(derivatives with respect to it will be denoted by primes),
$k$ the wave-number,
$H=a'/a^2$ the Hubble parameter and
\be
\Omega^2(k,\eta)\equiv k^2-\frac{z''}{z}
\ ,
\label{freq}
\ee
where $z=z_{\rm S}\equiv a^2\,\phi'/H$ for scalar
and $z=z_{\rm T}\equiv a$ for tensor perturbations
($\phi$ is the homogenous inflaton).
The dimensionless power spectra of scalar and tensor
fluctuations are then given by
\numparts
\be
\mathcal{P}_{\zeta}\equiv
\displaystyle\frac{k^{3}}{2\,\pi^{2}}\,
\left|\frac{\mu_{\rm S}}{z_{\rm S}}\right|^{2}
\ ,
\ \ \ \
\mathcal{P}_{h}\equiv
\displaystyle\frac{4\,k^{3}}{\pi^{2}}\,
\left|\frac{\mu_{\rm T}}{z_{\rm T}}\right|^{2}
\label{spectra_def}
\ee
and the spectral indices and runnings by
\be
&&
n_{\rm S}-1\equiv
\left.\displaystyle\frac{\d\ln \mathcal{P}_{\zeta}}
{\d\ln k}\right|_{k=k_{*}}
\ ,
\ \ \
n_{\rm T}\equiv
\left.\displaystyle\frac{\d\ln \mathcal{P}_{h}}
{\d\ln k}\right|_{k=k_{*}}
\label{n_def}
\\
&&
\alpha_{\rm S}\equiv\left.
\frac{\d^{2}\ln\mathcal{P}_{\zeta}}
{(\d\ln k)^{2}}\right|_{k=k_{*}}
\ ,
\ \ \
\alpha_{\rm T}\equiv\left.
\frac{\d^{2}\ln \mathcal{P}_{h}}
{(\d\ln k)^{2}}\right|_{k=k_{*}}
\label{alpha_def}
\ee
where $k_*$ is an arbitrary pivot scale.
We also define the tensor-to-scalar ratio at $k=k_*$ as
\be
R\equiv\left.\frac{\mathcal{P}_{h}}{\mathcal{P}_{\zeta}}
\right|_{k=k_{*}}
\ .
\label{R_def}
\ee
\endnumparts
Finally, in the following we shall often make use of the hierarchy of
horizon flow functions (HFF in short, also referred to as slow-roll
parameters) $\epsilon_i$'s defined by~\cite{terrero}
\be
\epsilon_1 \equiv -\frac{\dot{H}}{H^2}
\ ,
\quad\quad
\epsilon_{n+1} \equiv \frac{\dot{\epsilon}_n}{H\,\epsilon_n}
\quad n\ge1
\label{HFF_def}
\ee
where dots denote derivatives with respect to the cosmic time
$\d t=a\,\d\eta$.
\section{MCE and cosmological perturbations}
\label{sMCECP}
In order to apply the MCE to cosmological perturbations, we shall
start from the same equation as was used with the improved WKB
method in Refs.~\cite{MS,WKB1}, to which we refer for more details.
We recall here that the WKB approximation can be more effectively
applied {\em after\/} the following redefinitions of the wave-function
and variable,
\numparts
\be
&&
\chi=(1-\epsilon_1)^{1/2}\,{\rm e}^{-x/2}\,\mu
\\
&&
x=\ln\left(\frac{k}{H\,a}\right)
\ .
\ee
\endnumparts
This yields an equation of the form~(\ref{exact_EQ}) with
the ``frequency'' $\Omega(k,\eta)$ of Eq.~(\ref{osci}) replaced by
\be
\omega^2(x)=\frac{{\rm e}^{2\,x}}{\left[1-\epsilon_1(x)\right]^2}-\nu^2(x)
\ ,
\label{our_freq}
\ee
with $\nu^2(x)$ given, respectively for scalar and tensor
perturbations, by
\numparts
\be
\!\!\!\!\!\!\!\!\!\!\!\!\!\!\!\!\!\!\!\!\!\!\!\!\!
\nu_{\rm S}^2(x)&=&
\frac{1}{4}\,\left(\frac{3-\epsilon_1}{1-\epsilon_1}\right)^2
+\frac{(3-2\,\epsilon_1)\,\epsilon_2}{2\,(1-\epsilon_1)^2}
+\frac{(1-2\,\epsilon_1)\,\epsilon_2\,\epsilon_3}{2\,(1-\epsilon_1)^3}
+\frac{(1-4\,\epsilon_1)\,\epsilon_2^2}{4\,(1-\epsilon_1)^4}
\label{nu2_S}
\\
\!\!\!\!\!\!\!\!\!\!\!\!\!\!\!\!\!\!\!\!\!\!\!\!\!
\nu_{\rm T}^2(x)&=&
\frac{1}{4}\,\left(\frac{3-\epsilon_1}{1-\epsilon_1}\right)^2
-\frac{\epsilon_1\,\epsilon_2}{2\,(1-\epsilon_1)^2}
-\frac{\epsilon_1\,\epsilon_2\,\epsilon_3}{2\,(1-\epsilon_1)^3}
-\frac{(2+\epsilon_1)\,\epsilon_1\,\epsilon_2^2}{4\,(1-\epsilon_1)^4}
\ ,
\label{nu2_T}
\ee
\endnumparts
where we omit the dependence on $x$ in the $\epsilon_i$ for the sake of
brevity.
The point $x=x_0$ where the frequency vanishes, $\omega(x_0)=0$
(i.e.~the classical ``turning point''), is given by the expression
\be
x_0=\ln\left[\bar{\nu}\,\left(1-\bar{\epsilon}_1\right)\right]
\ ,
\ee
where we have defined $\bar{\nu}\equiv\nu(x_0)$ and
$\bar{\epsilon}_1\equiv\epsilon_1(x_0)$.
We now choose the comparison function
\be
\Theta^2(\sigma)=
\frac{{\rm e}^{2\,\sigma}}{(1-\bar{\epsilon}_1)^2}-\bar{\nu}^2
\ ,
\label{aux_FREQ}
\ee
and note that then $\sigma_0=x_0$.
Solutions to Eq.~(\ref{exact_EQ}) can now be expressed, by means of
Eqs.~(\ref{aux_EQ}), (\ref{new_EQ_appr}) and (\ref{aux_FREQ}), as
\be
\chi_{\pm}(x)\simeq\sqrt{\frac{\Theta(\sigma)}{\omega(x)}}\,
J_{\pm\bar{\nu}}\left(\frac{{\rm e}^{\sigma}}{1-\bar{\epsilon}_1}\right)
\ ,
\label{exact_SOL_bis}
\ee
where the $J$'s are Bessel functions~\cite{abram}, and the
initial condition~(\ref{init_cond_on_mu}) can be satisfied by taking
a linear combination of them.
However, in contrast with the WKB method~\cite{MS,WKB1} and as we pointed
out in Section~\ref{sMCE}, MCE solutions need not be matched at the turning 
point, since the functions~(\ref{exact_SOL_bis}) are valid solutions
for the whole range of the variable $x$.
Eq.~(\ref{new_EQ_int}) at the end of inflation, $x=x_{\rm f}$, becomes
\be
\xi(x_{\rm f})
&\simeq&
-\Theta(\sigma_{\rm f})
-\frac{\bar{\nu}}{2}\,\ln\left[
\frac{\bar{\nu}-\Theta(\sigma_{\rm f})}
{\bar{\nu}+\Theta(\sigma_{\rm f})}\right]
\nonumber
\\
&\simeq&
-\bar{\nu}\,
\left[1+\ln\left(\frac{{\rm e}^{\sigma_{\rm f}}}{1-\bar{\epsilon}_1}\right)
-\ln\left(2\,\bar{\nu}\right)\right]
\ ,
\label{new_EQ_integrate}
\ee
where the super-horizon limit
$x_{\rm f}\ll x_0$ ($\sigma_{\rm f}\to-\infty$)
has been taken in the second line.
One then has
\be
\frac{{\rm e}^{\sigma_{\rm f}}}{1-\bar{\epsilon}_1}
\simeq
\frac{2\,\bar{\nu}}{\rm e}\,{\rm exp}
\left[-\frac{\xi(x_{\rm f})}{\bar{\nu}}\right]
\ .
\label{arg_exp}
\ee
Finally, on using Eq.~(\ref{arg_exp}), we obtain the general expressions
for the power spectra to leading MCE order,
\numparts
\be
{\cal P}_\zeta
&=&
\left[\frac{H^2}{\pi\,\epsilon_1\,m_{\rm Pl}^2}
\left(\frac{k}{a\,H}\right)^3
\frac{{\rm e}^{2\,\xi_{\rm S}}}
{\left(1-\epsilon_1\right)\,\omega_{\rm S}}\right]_{x=x_{\rm f}}
g_{\rm S}(x_0)
\label{spectra_S}
\\
{\cal P}_h
&=&
\left[\frac{16\,H^2}{\pi\,m_{\rm Pl}^2}\,
\left(\frac{k}{a\,H}\right)^3\,
\frac{{\rm e}^{2\,\xi_{\rm T}}}
{\left(1-\epsilon_1\right)\,\omega_{\rm T}}\right]_{x=x_{\rm f}}
g_{\rm T}(x_0)
\label{spectra_T}
\ ,
\ee
where $m_{\rm Pl}$ is the Planck mass and the quantities inside the
square brackets are evaluated in the super-horizon limit, whereas
the functions
\be
g(x_0)
\equiv
\frac{\pi\,e^{2\,\bar{\nu}}\,\bar{\nu}^{1-2\,\bar{\nu}}}
{\left[1-\cos\left(2\,\pi\,\bar{\nu}\right)\right]\,
\left[\Gamma\left(1-\bar{\nu}\right)\right]^2}
\ ,
\label{corr_TP}
\ee
\endnumparts
describe corrections that just depend on quantities evaluated
at the turning point and represent the main result of the
MCE applied to cosmological perturbations~\footnote{Inside the
square brackets we recognize the general results given by the WKB
approximation~\cite{MS,WKB1}.
The ``correction'' $g(x_0)$ accounts for the fact that $\Theta^2$
in Eq.~(\ref{aux_FREQ}) is a better approximation than
Langer's~\cite{WKB1,langer}.}.
The expression in Eq.~(\ref{corr_TP}) is obtained by simply making use of
the approximate solutions~(\ref{exact_SOL_bis}) and their asymptotic
expansion at $x\to\infty$ to impose the initial
conditions~(\ref{init_cond_on_mu}).
In the WKB calculations~\cite{WKB_PLB,WKB_lungo,WKB1} one finds a similar
factor but, in that case, using the Bessel functions of order $1/3$ leads
to a large error in the amplitudes.
The MCE instead uses Bessel functions of order $\bar{\nu}$, with $\bar\nu=3/2$
to leading order in the HFF (i.e.~the right index for the de~Sitter model),
which yields a significantly better value for the amplitudes of inflationary
spectra.
\par
The MCE allows one to compute approximate perturbation modes with errors
in the asymptotic regions (i.e.~in the sub- and super-horizon limits)
which are comparable with those of the standard (or improved) WKB
approximation~\cite{MS,WKB1}.
Since these methods usually give large errors at the turning
point~\cite{WKB1} (which produce equally large errors in
the amplitude of the power spectra) it will suffice to estimate the
error at the turning point in order to show that the MCE is indeed
an improvement.
To leading order (that is, on using the approximate
solution~(\ref{exact_SOL_bis})), the MCE gives an error at the
turning point of the second order in the HFF, which means that we
have a small error in the amplitudes of the power spectra.
Unfortunately, this error remains of second order
in the HFF also for next-to-leading order in the MCE.
We shall see this by applying Dingle's analysis~\cite{dingle}
for linear frequencies to our case~(\ref{our_freq}).
We start by rewriting Eq.~(\ref{new_EQ}) as
\be
\left\{\omega^2-\sigma_1^2\,\left[\frac{{\rm e}^{2\,\sigma}}
{(1-\bar{\epsilon}_1)^2}-\bar{\nu}^2\right]
\right\}
+
\left[
\frac34\,\frac{\sigma_2^2}{\sigma_1^2}-\frac12\,\frac{\sigma_3}{\sigma_1}
\right]
=0
\ ,
\label{new_EQ_4_DING}
\ee
where we dropped the $x$ dependence in $\omega$ and $\sigma$
and the order of the derivatives is given by their subscripts
($\sigma_1\equiv \d\sigma/\d x$, $\omega_1^2\equiv \d\omega^2/\d x$,
etc.).
Note that the term in square brackets is the error obtained on using
the solutions~(\ref{exact_SOL_bis}).
We then evaluate Eq.~(\ref{new_EQ_4_DING}) and its subsequent
derivatives at the turning point (i.e.~at $x=x_0$),
\numparts
\be
\!\!\!\!\!\!\!\!\!\!\!\!\!\!\!\!\!\!\!\!\!\!\!\!\!\!\!\!\!\!\!\!\!\!\!\!\!\!\!\!
\left\{
\omega^2-\sigma_1^2\,\Theta^2\left(\sigma\right)
\right\}
+
\left[
\frac34\,\frac{\sigma_2^2}{\sigma_1^2}-\frac12\,\frac{\sigma_3}{\sigma_1}
\right]
=0
\label{new_EQ_4_DING_TP}
\ee
\be
\!\!\!\!\!\!\!\!\!\!\!\!\!\!\!\!\!\!\!\!\!\!\!\!\!\!\!\!\!\!\!\!\!\!\!\!\!\!\!\!
\left\{
\omega^2_1
-2\,\sigma_2\,\sigma_1\,\Theta^2\left(\sigma\right)
-\frac{2\,{\rm e}^{2\,\sigma}\,\sigma_1^3}{\left(1-\bar{\epsilon}_1\right)^2}
\right\}
+
\left[
2\,\frac{\sigma_3\,\sigma_2}{\sigma_1^2}
-\frac32\,\frac{\sigma_2^3}{\sigma_1^3}
-\frac12\,\frac{\sigma_4}{\sigma_1}
\right]
=0
\label{new_EQ_deriv1}
\ee
\be
\!\!\!\!\!\!\!\!\!\!\!\!\!\!\!\!\!\!\!\!\!\!\!\!\!\!\!\!\!\!\!\!\!\!\!\!\!\!\!\!
&&
\left\{
\omega^2_2
-2\,\left(\sigma_3\,\sigma_1+\sigma_2^2\right)\,\Theta^2\left(\sigma\right)
-\frac{2\,\sigma_1^2\,{\rm e}^{2\,\sigma}}{\left(1-\bar{\epsilon}_1\right)^2}\,
\left(2\,\sigma_1^2+5\,\sigma_2\right)
\right\}
\nonumber\\
\!\!\!\!\!\!\!\!\!\!\!\!\!\!\!\!\!\!\!\!\!\!\!\!\!\!\!\!\!\!\!\!\!\!\!\!\!\!\!\!
&&
+
\left[
\frac92\,\frac{\sigma_2^4}{\sigma_1^4}
-\frac{17}{2}\frac{\sigma_3\,\sigma_2^2}{\sigma_1^3}
+\frac52\,\frac{\sigma_4\,\sigma_2}{\sigma_1^2}
+2\,\frac{\sigma_3^2}{\sigma_1^2}
-\frac12\,\frac{\sigma_5}{\sigma_1}
\right]
=0
\label{new_EQ_deriv2}
\ee
\be
\!\!\!\!\!\!\!\!\!\!\!\!\!\!\!\!\!\!\!\!\!\!\!\!\!\!\!\!\!\!\!\!\!\!\!\!\!\!\!\!
&&
\left\{
\omega^2_3
-2\,\left(\sigma_4\,\sigma_1+3\,\sigma_2\,\sigma_3\right)\,\Theta^2\left(\sigma\right)
-\frac{2\,\sigma_1\,{\rm e}^{2\,\sigma}}{\left(1-\bar{\epsilon}_1\right)^2}\,
\left(4\,\sigma_1^4
+18\,\sigma_2\,\sigma_1^2
+7\,\sigma_3\,\sigma_1
+12\,\sigma_2^2\right)
\right\}
\nonumber
\\
\!\!\!\!\!\!\!\!\!\!\!\!\!\!\!\!\!\!\!\!\!\!\!\!\!\!\!\!\!\!\!\!\!\!\!\!\!\!\!\!
&&
+\left[
\frac{87}{2}\,\frac{\sigma_3\,\sigma_2^3}{\sigma_1^4}
-18\,\frac{\sigma_2^5}{\sigma_1^5}
-\frac{27}{2}\,\frac{\sigma_4\,\sigma_2^2}{\sigma_1^3}
-21\,\frac{\sigma_3^2\,\sigma_2}{\sigma_1^3}
+3\,\frac{\sigma_5\,\sigma_2}{\sigma_1^2}
+\frac{13}{2}\,\frac{\sigma_3\,\sigma_4}{\sigma_1^2}
-\frac12\,\frac{\sigma_6}{\sigma_1}
\right]
=0
\ ,
\label{new_EQ_deriv3}
\ee
\endnumparts
where $\Theta^2\left(\sigma\right)$ was defined in Eq.~(\ref{aux_FREQ})
and we omit two equations for brevity.
In order to evaluate the error at the turning point
\be
\Delta_{\rm TP}=
\left[\frac34\,\frac{\sigma_2^2}{\sigma_1^2}
-\frac12\,\frac{\sigma_3}{\sigma_1}\right]_{x=x_0}
\ ,
\ee
we ignore the terms in square brackets and equate to zero the
expressions in the curly brackets in
Eqs.~(\ref{new_EQ_4_DING_TP})-(\ref{new_EQ_deriv3}) and so on.
This leads to
\numparts
\be
\!\!\!\!\!\!\!\!\!\!\!\!\!\!\!\!\!\!\!\!\!\!\!\!\!\!\!\!\!\!\!\!\!\!\!\!\!\!
\sigma
&\!=\!&
\ln\left[\left(1-\bar{\epsilon}_1\right)\,\bar{\nu}\right]
\label{sigma0_0order}
\\
\!\!\!\!\!\!\!\!\!\!\!\!\!\!\!\!\!\!\!\!\!\!\!\!\!\!\!\!\!\!\!\!\!\!\!\!\!\!
\sigma_1
&\!=\!&
\left(\frac{\omega^2_1}{2\,\bar{\nu}^{2}}\right)^{1/3}
\label{sigma1_0order}
\\
\!\!\!\!\!\!\!\!\!\!\!\!\!\!\!\!\!\!\!\!\!\!\!\!\!\!\!\!\!\!\!\!\!\!\!\!\!\!
\sigma_2
&\!=\!&
\frac{1}{5\,\left(2\,\bar{\nu}^2\right)^{1/3}}
\left[
\frac{\omega^2_2}{\left(\omega^2_1\right)^{2/3}}
-\left(2\,\frac{\omega^2_1}{\bar{\nu}}\right)^{2/3}
\right]
\label{sigma2_0order}
\\
\!\!\!\!\!\!\!\!\!\!\!\!\!\!\!\!\!\!\!\!\!\!\!\!\!\!\!\!\!\!\!\!\!\!\!\!\!\!
\sigma_3
&\!=\!&
-\frac{6 \left(2^{1/3}\,\omega^2_2\right)^2}
{175\,\left(\bar{\nu}^{2/5}\,\omega^2_1\right)^{5/3}}
-\frac{3\cdot 2^{1/3}\,\omega^2_2}
{25\,\left(\bar{\nu}^{4}\,\omega^2_1\right)^{1/3}}
+\frac{16\,\omega^2_1}{175\,\bar{\nu}^2}
+\frac{\omega^2_3}
{7 \left(2^{1/2}\,\bar{\nu}\,\omega^2_1\right)^{2/3}}
\ ,
\label{sigma3_0order}
\ee
\endnumparts
and similar expressions for $\sigma_4$, $\sigma_5$, and $\sigma_6$
which we again omit for brevity.
On inserting Eqs.~(\ref{our_freq}), (\ref{nu2_S}) and (\ref{nu2_T})
in the above expressions, we find the errors to leading MCE order
\numparts
\be
\Delta^{(0)}_{\rm TP, S}&=&
-\frac{32}{315}\,\epsilon_1\,\epsilon_2
-\frac{22}{315}\,\epsilon_2\,\epsilon_3
\label{corr_TP_S_0order}
\\
\Delta^{(0)}_{\rm TP, T}&=&
-\frac{32}{315}\,\epsilon_1\,\epsilon_2
\ ,
\label{corr_TP_T_0order}
\ee
\endnumparts
for scalar and tensor modes respectively.
\par
On iterating this procedure we can further obtain the errors
for the next-to-leading MCE order $\Delta^{(1)}_{\rm TP}$.
We first compute next-to-leading solutions to
Eqs.~(\ref{new_EQ_4_DING_TP})-(\ref{new_EQ_deriv3}) and so on
by inserting the solutions found to leading order for
$\sigma_1,\sigma_2,\ldots,\sigma_6$ into the corrections
(i.e.~the square brackets) and into all terms containing
$\Theta^2(\sigma)$~\cite{dingle}.
This leads to
\numparts
\be
\Delta^{(1)}_{\rm TP, S}&=&
-\frac{31712}{331695}\,\epsilon_1\,\epsilon_2
-\frac{21598}{331695}\,\epsilon_2\,\epsilon_3
\label{corr_TP_S_1order}
\\
\Delta^{(1)}_{\rm TP, T}&=&
-\frac{31712}{331695}\,\epsilon_1\,\epsilon_2
\ ,
\label{corr_TP_T_1order}
\ee
\endnumparts
which show that the next-to-leading MCE solutions lead to an error of
second order in the HFF, too.
We suspect that this remains true for higher MCE orders, since there
is no {\em a priori\/} relation between the MCE and the
slow-roll expansions.
Let us however point out that the above expressions were obtained
without performing a slow-roll expansion and therefore do not
require that the $\epsilon_i$ be small.
\section{Applications}
\label{s_app}
In this section we apply the formalism developed in the previous section
to some models of inflation.
We shall expand our general expressions (\ref{spectra_S})-(\ref{corr_TP})
to second order in the HFF and compare them with other approximation methods used
in the literature.
\subsection{Power-law~inflation}
\label{power-law}
In this model~\cite{PL,LS}, the scale factor is given
in conformal time by
\be
a(\eta)=\ell_0\left|\eta\right|^{1+\beta}
\ ,
\label{a}
\ee
where $\beta \le -2$ and $\ell_0=H^{-1}$ corresponds to the
(constant) Hubble radius for de~Sitter ($\beta=-2$).
Since the HFF are constant,
\be
\epsilon_1 = \frac{2+\beta}{1+\beta}
\ ,
\quad \quad
\epsilon_n = 0
\ ,
\quad
n>1
\ ,
\label{eps_PL}
\ee
the MCE yields the exact power spectra, spectral indices and runnings,
\numparts
\be
{\cal P}_{\zeta}=
\frac{\ell_{\rm Pl}^2}{\ell_0^2\,\pi\,\epsilon_1}\,f(\beta)\,k^{2\beta+4}
\label{PL_spectra_S}
\\
{\cal P}_h=
\frac{16\,\ell_{\rm Pl}^2}{\ell_0^2\,\pi}\,f(\beta)k^{2\beta+4}
\label{PL_spectra_T}
\ee
where $\ell_{\rm Pl}=m_{\rm Pl}^{-1}$ is the Planck length and
\be
\!\!\!\!\!\!\!\!\!\!\!\!\!\!\!\!\!\!\!\!\!\!\!\!\!\!\!\!\!\!
f(\beta)=\frac{\pi}{2^{2\,\beta+1}}\,
\frac{1}{\left[1-\cos\left(2\,\pi\left|\beta+\frac12\right|\right)\right]\,
\Gamma^2\left(\beta+\frac32\right)}
\equiv
\frac{1}{\pi}\,
\left[\frac{\Gamma\left(\left|\beta+\frac12\right|\right)}{2^{\beta+1}}\right]^2
\ ,
\label{f}
\ee
\endnumparts
with $\Gamma$ the Gamma function.
The spectral indices are $n_{\rm S}-1=n_{\rm T}=2\beta+4$ and
their runnings $\alpha_{\rm S}=\alpha_{\rm T}=0$.
Finally, the tensor-to-scalar ratio becomes
\be
R=16\,\frac{2+\beta}{1+\beta}
\ ,
\label{R_PL}
\ee
which is constant as well.
\subsection{Leading~MCE and second~slow-roll~order}
\label{lead_MCE_and_2SR}
We now consider the results~(\ref{spectra_S})-(\ref{corr_TP}) given by
the MCE to leading order (denoted by the subscript MCE) and evaluate them
to second~order in the HFF (labelled by the superscript $(2)$) for a general
inflationary scale factor.
A crucial point in our method is the computation of the function $\xi$
defined in Eq.~(\ref{new_EQ_int}) which can be found in detail in Section~III
of Ref.~\cite{WKB_lungo}.
For the sake of brevity, we shall not reproduce that analysis here but
it is important to stress that, in contrast with the GFM and other slow-roll
approximations, it does not require {\em a priori\/} any expansion in the
HFF since Eq.~(34) of Ref.~\cite{WKB_lungo} is exact and higher~order terms
are discarded {\em a fortiori\/}.
From that expression, upon neglecting terms of order higher than two in the HFF,
we obtain the power spectra
\numparts
\be
\!\!\!\!\!\!\!\!\!\!\!\!\!\!\!\!\!\!\!\!\!\!\!\!\!\!\!\!\!\!\!\!\!\!\!\!\!\!\!\!\!\!\!
&&
\mathcal{P}_{\zeta,\scriptscriptstyle\scriptscriptstyle{\rm MCE}}^{(2)}
\!\!=\!\!
\frac{H^2}{\pi\epsilon_1m_{\rm Pl}^2}\!
\left\{1\!-\!2\left(C\!+\!1\right)\epsilon_1\!-\!C\,\epsilon_2
\!+\!\left(2C^2\!+\!2C\!+\!\frac{\pi^2}{2}\!-\!5\right)\epsilon_1^2
\!+\!\left(\frac12C^2\!+\!\frac{\pi^2}{8}\!-\!1\right)\epsilon_2^2
\right.
\nonumber
\\
\!\!\!\!\!\!\!\!\!\!\!\!\!\!\!\!\!\!\!\!\!\!\!\!\!\!\!\!\!\!\!\!\!\!\!\!\!\!\!\!\!\!\!
&&
\,\,\,\,\,\,
+\!\!
\left.
\left(2\,C^2\!-\!2\,C\,D_{\scriptscriptstyle{\rm MCE}}
\!+\!D_{\scriptscriptstyle{\rm MCE}}^2\!-\!C\!-\!2\,C\,\ln(2)
\!+\!2\,D_{\scriptscriptstyle{\rm MCE}}\,\ln(2)
\!+\!\frac{7\pi^2}{12}\!-\!\frac{64}{9}\right)\epsilon_1\,\epsilon_2
\right.
\nonumber
\\
\!\!\!\!\!\!\!\!\!\!\!\!\!\!\!\!\!\!\!\!\!\!\!\!\!\!\!\!\!\!\!\!\!\!\!\!\!\!\!\!\!\!\!
&&
\,\,\,\,\,\,
+\!\!
\left.
\left(-C\,D_{\scriptscriptstyle{\rm MCE}}
\!+\!\frac12\,D_{\scriptscriptstyle{\rm MCE}}^2\!-\!C\,\ln(2)
\!+\!D_{\scriptscriptstyle{\rm MCE}}\,\ln(2)\!+\!\frac{\pi^2}{24}
\!-\!\frac{1}{18}\right)\epsilon_2\,\epsilon_3
\right.
\nonumber
\\
\!\!\!\!\!\!\!\!\!\!\!\!\!\!\!\!\!\!\!\!\!\!\!\!\!\!\!\!\!\!\!\!\!\!\!\!\!\!\!\!\!\!\!
&&
\,\,\,\,\,\,
+\!\!\left.
\left[-2\,\epsilon_1\!-\!\epsilon_2\!+\!2\left(2\,C\!+\!1\right)
\epsilon_1^2
\!+\!\left(4\,C\!-\!2\,D_{\scriptscriptstyle{\rm MCE}}\!-\!1\right)\epsilon_1\,\epsilon_2
\!+\!C\,\epsilon_2^2\!-\!D_{\scriptscriptstyle{\rm MCE}}\,\epsilon_2\,\epsilon_3\right]
\ln\left(\frac{k}{k_*}\right)
\right.
\nonumber
\\
\!\!\!\!\!\!\!\!\!\!\!\!\!\!\!\!\!\!\!\!\!\!\!\!\!\!\!\!\!\!\!\!\!\!\!\!\!\!\!\!\!\!\!
&&
\,\,\,\,\,\,
+\!\!
\left.
\frac12\left(4\,\epsilon_1^2\!
+\!2\,\epsilon_1\,\epsilon_2\!
+\!\epsilon_2^2
\!-\!\epsilon_2\,\epsilon_3\right)
\ln^2\left(\frac{k}{k_*}\right)\right\}
\label{PS_SlowRoll_0_2order}
\ee
\be
\!\!\!\!\!\!\!\!\!\!\!\!\!\!\!\!\!\!\!\!\!\!\!\!\!\!\!\!\!\!\!\!\!\!\!\!\!\!\!\!\!\!\!
\mathcal{P}_{h,\scriptscriptstyle{\rm MCE}}^{(2)}
&\!=\!&
\frac{16H^2}{\pi m_{\rm Pl}^2}
\left\{1-2\left(C+1\right)\epsilon_1
+\left(2C^2+2C+\frac{\pi^2}{2}-5\right)\epsilon_1^2
\nonumber
\right.
\\
\!\!\!\!\!\!\!\!\!\!\!\!\!\!\!\!\!\!\!\!\!\!\!\!\!\!\!\!\!\!\!\!\!\!\!\!\!\!\!\!\!\!\!
&&
+\!\!
\left.
\left(-2CD_{\scriptscriptstyle{\rm MCE}}+D_{\scriptscriptstyle{\rm MCE}}^2
-2C-2C\ln(2)+2D_{\scriptscriptstyle{\rm MCE}}\ln(2)
+\frac{\pi^2}{12}-\frac{19}{9}\right)\epsilon_1\epsilon_2
\nonumber
\right.
\\
\!\!\!\!\!\!\!\!\!\!\!\!\!\!\!\!\!\!\!\!\!\!\!\!\!\!\!\!\!\!\!\!\!\!\!\!\!\!\!\!\!\!\!
&&
+\!\!
\left.
\left[-2\epsilon_1+2\left(2C+1\right)\epsilon_1^2
-2\left(D_{\scriptscriptstyle{\rm MCE}}+1\right)\epsilon_1\epsilon_2\right]\,
\ln\left(\frac{k}{k_*}\right)
\nonumber
\right.
\\
\!\!\!\!\!\!\!\!\!\!\!\!\!\!\!\!\!\!\!\!\!\!\!\!\!\!\!\!\!\!\!\!\!\!\!\!\!\!\!\!\!\!\!
&&
+\!\!
\left.\frac12\left(4\epsilon_1^2
-2\epsilon_1\epsilon_2\right)\ln^2\left(\frac{k}{k_*}\right)
\right\}
\ ,
\label{PT_SlowRoll_0_2order}
\ee
\endnumparts
where $D_{\scriptscriptstyle{\rm MCE}}\equiv\frac{1}{3}-\ln 3\approx -0.7652$
and $C\equiv\ln 2+\gamma_{\rm E}-2\approx -0.7296$, with $\gamma_{\rm E}$
the Euler-Mascheroni constant.
A clarification concerning the $g(x_0)$'s is in order.
Since the turning point does not coincide with the horizon crossing
where the spectra are evaluated~\cite{WKB_lungo},
we have used the relation
\be
\epsilon_i(x_0)
\simeq
\epsilon_i-\epsilon_i\,\epsilon_{i+1}\,\ln\left(\frac32\right)
\ ,
\label{epsilon_n_TPtoHC}
\ee
in order to express the $g(x_0)$'s as functions of the crossing time
(HFF with no explicit argument are evaluated at the horizon crossing).
The spectral indices~(\ref{n_def}) are then given by
\numparts
\be
&&
\!\!\!\!\!\!\!\!\!\!\!
n_{\rm S,\scriptscriptstyle{\rm MCE}}^{(2)}-1
=
-2\,\epsilon_1-\epsilon_2-2\,\epsilon_1^2
-\left(2\,D_{\scriptscriptstyle{\rm MCE}}+3\right)\,\epsilon_1\,\epsilon_2
-D_{\scriptscriptstyle{\rm MCE}}\,\epsilon_2\,\epsilon_3
\label{n_2'order_S}
\\
&&
\!\!\!\!\!\!\!\!\!\!\!
n_{\rm T,\scriptscriptstyle{\rm MCE}}^{(2)}
=
-2\,\epsilon_1-2\,\epsilon_1^2
-2\,\left(D_{\scriptscriptstyle{\rm MCE}}+1\right)\,\epsilon_1\,\epsilon_2
\ ,
\label{n_2'order_T}
\ee
and their runnings~(\ref{alpha_def}) by
\be
&&
\alpha_{\rm S,\scriptscriptstyle{\rm MCE}}^{(2)}=
-2\,\epsilon_1\,\epsilon_2
-\epsilon_2\,\epsilon_3
\label{alpha_2'order_S}
\\
&&
\alpha_{\rm T,\scriptscriptstyle{\rm MCE}}^{(2)}=
-2\,\epsilon_1\,\epsilon_2
\label{alpha_2'order_T}
\ .
\ee
\endnumparts
The tensor-to-scalar ratio~(\ref{R_def}) becomes
\be
\!\!\!\!\!\!\!\!\!\!\!\!\!\!\!\!\!\!\!\!\!\!\!\!\!\!\!\!\!\!\!\!\!\!\!\!
\frac{R^{(2)}_{\scriptscriptstyle{\rm MCE}}}{16\,\epsilon_1}
&=&
1+C\,\epsilon_2
+\left(C-\frac{\pi^2}{2}+5\right)
\epsilon_1\,\epsilon_2
+\left(\frac{1}{2}\,C^2-\frac{\pi^2}{8}+1\right)
\epsilon_2^2
\nonumber
\\
\!\!\!\!\!\!\!\!\!\!\!\!\!\!\!\!\!\!\!\!\!\!\!\!\!\!\!\!\!\!\!\!\!\!\!\!
&&
+\left(C\,D_{\scriptscriptstyle{\rm MCE}}
-\frac{1}{2}\,D_{\scriptscriptstyle{\rm MCE}}^2
+C\,\ln(2)
-D_{\scriptscriptstyle{\rm MCE}}\,\ln(2)
-\frac{\pi^2}{24}+\frac{1}{18}\right)
\epsilon_2\,\epsilon_3
\ .
\label{R_lead_WKB}
\ee
The polynomial structure in the HFF of the results agrees with
that given by the GFM~\cite{gongstewart,LLMS} and the
WKB~approximation~\cite{MS,WKB1,WKB_PLB,WKB_lungo} (the same polynomial
structure is also found for the spectral indices by means of the
uniform approximation~\cite{LA}).
Let us also note other aspects of our results:
first of all, the factors $g(x_0)$ modify the standard
WKB leading order amplitudes~\cite{MS,WKB1} so as to
reproduce the standard first order slow-roll results;
secondly, we found that $C$ and $D_{\scriptscriptstyle{\rm MCE}}$
are ``mixed'' in the numerical factors in front of second order
terms (we recall that $D_{\scriptscriptstyle{\rm MCE}}$ differs from
$C$ by about $5\,\%$); further, $D_{\scriptscriptstyle{\rm 
MCE}}=D_{\scriptscriptstyle{\rm WKB}}$
of Refs.~\cite{MS,WKB1,WKB_PLB,WKB_lungo}.
The runnings $\alpha_{\rm S}$ and $\alpha_{\rm T}$ are predicted to
be ${\mathcal O}(\epsilon^2)$~\cite{KT}, and in agreement with those
obtained by the GFM~\cite{gongstewart,LLMS}.
\begin{figure}[!ht]
\includegraphics[width=0.5\textwidth]{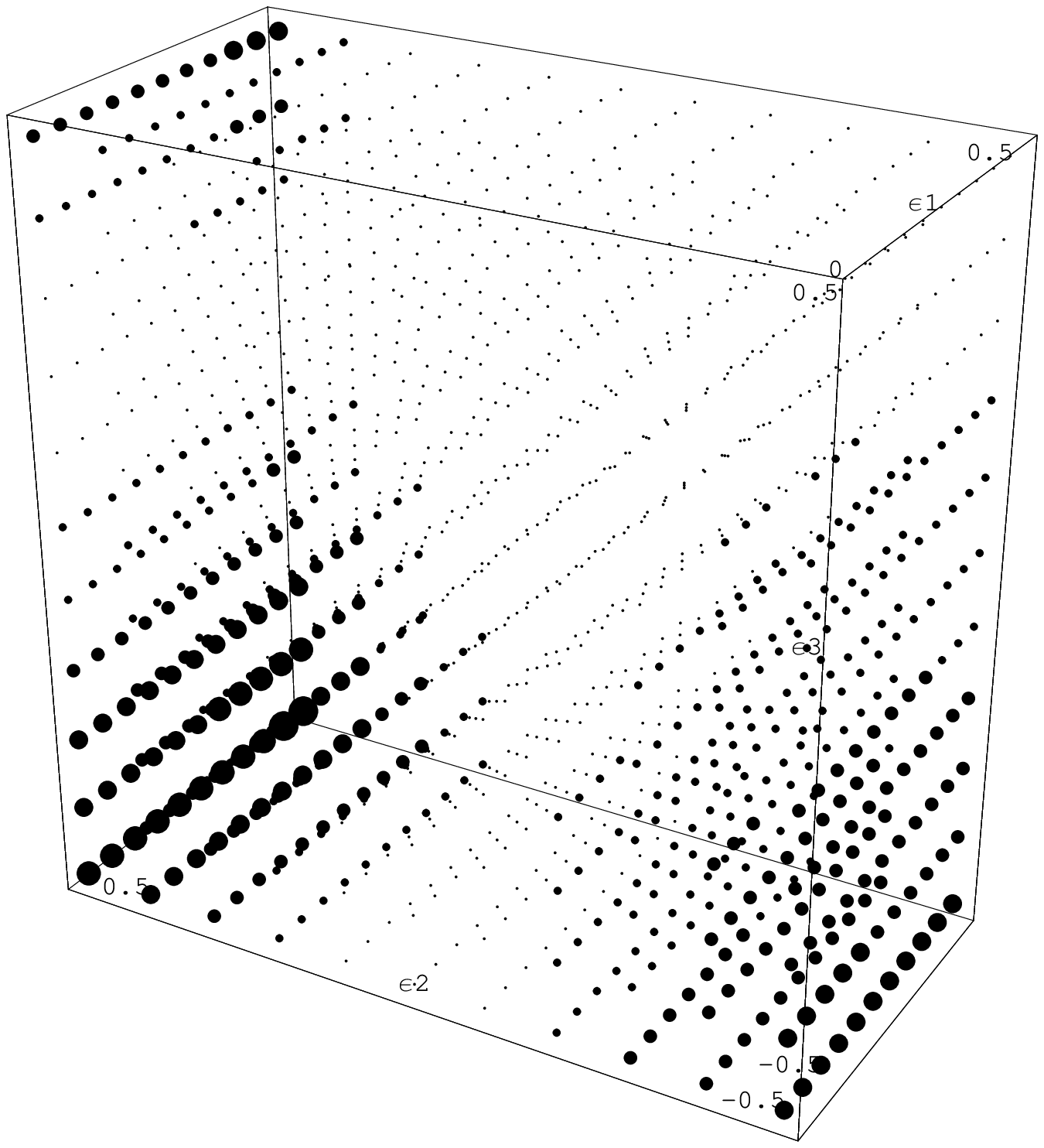}
\includegraphics[width=0.5\textwidth]{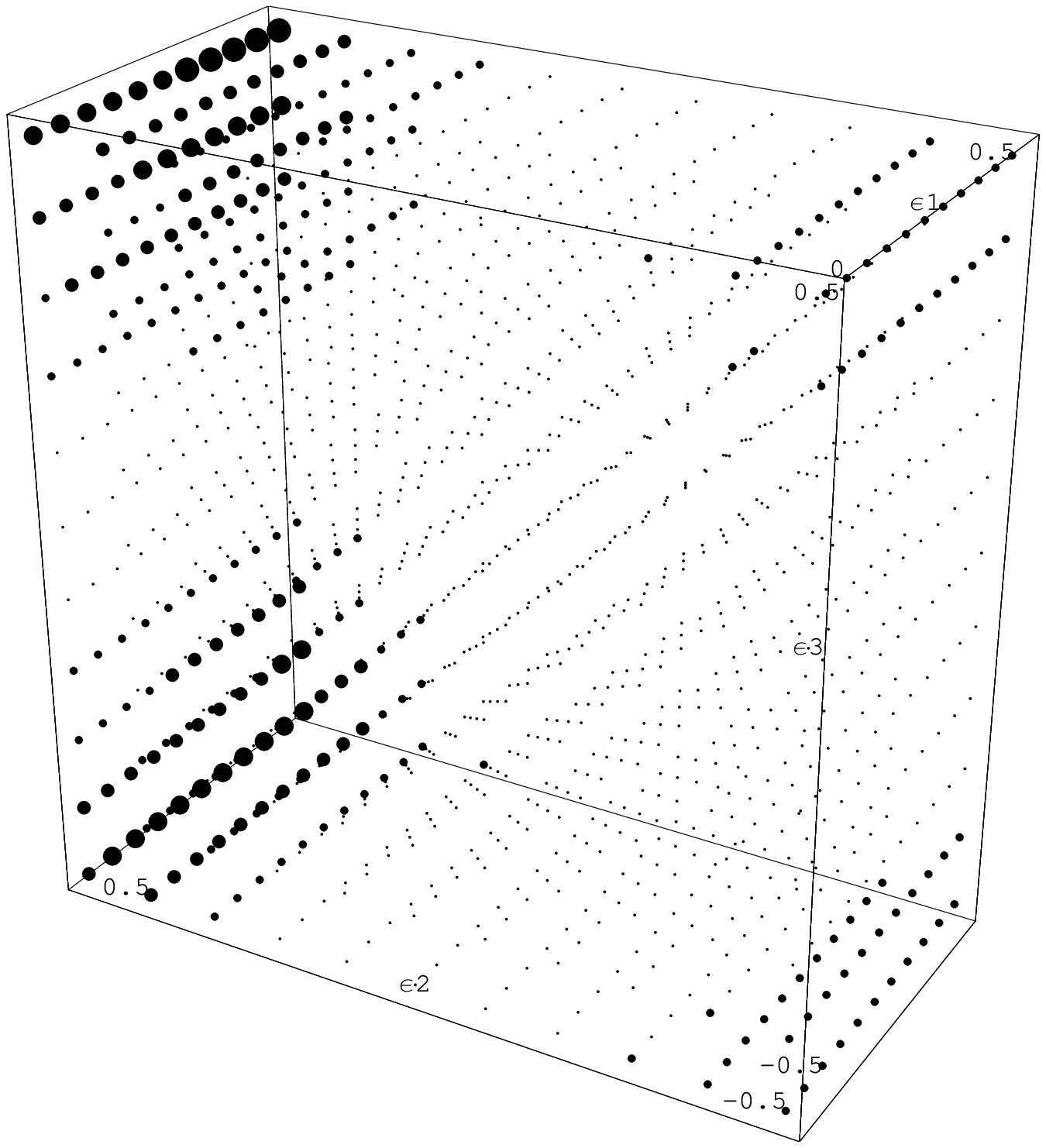}
\\
\null
\hspace{2cm}$\Delta R_{_{\rm WKB}}$
\hspace{6cm}$\Delta R_{_{\rm WKB*}}$
\\
\includegraphics[width=0.5\textwidth]{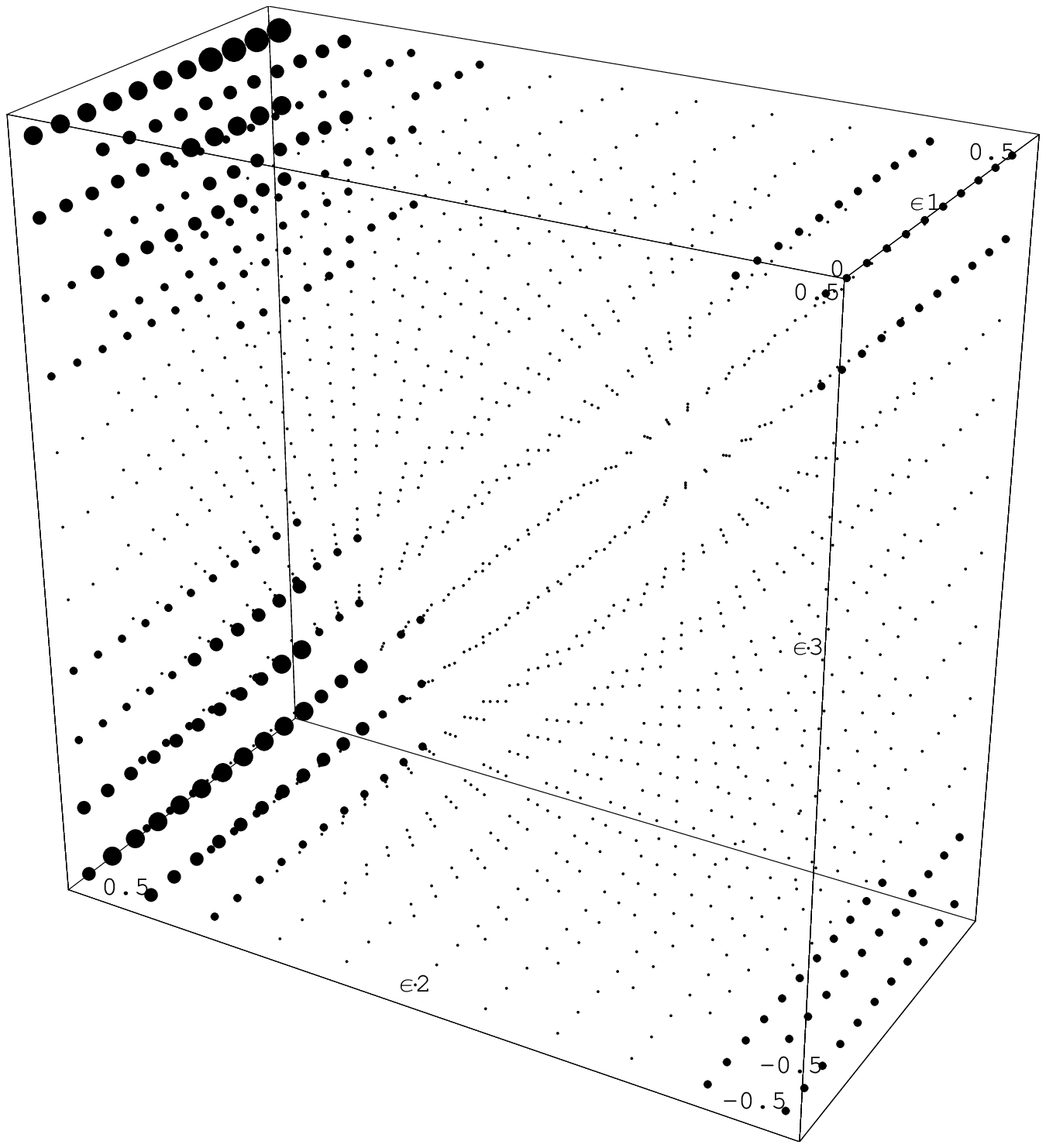}
\includegraphics[width=0.45\textwidth]{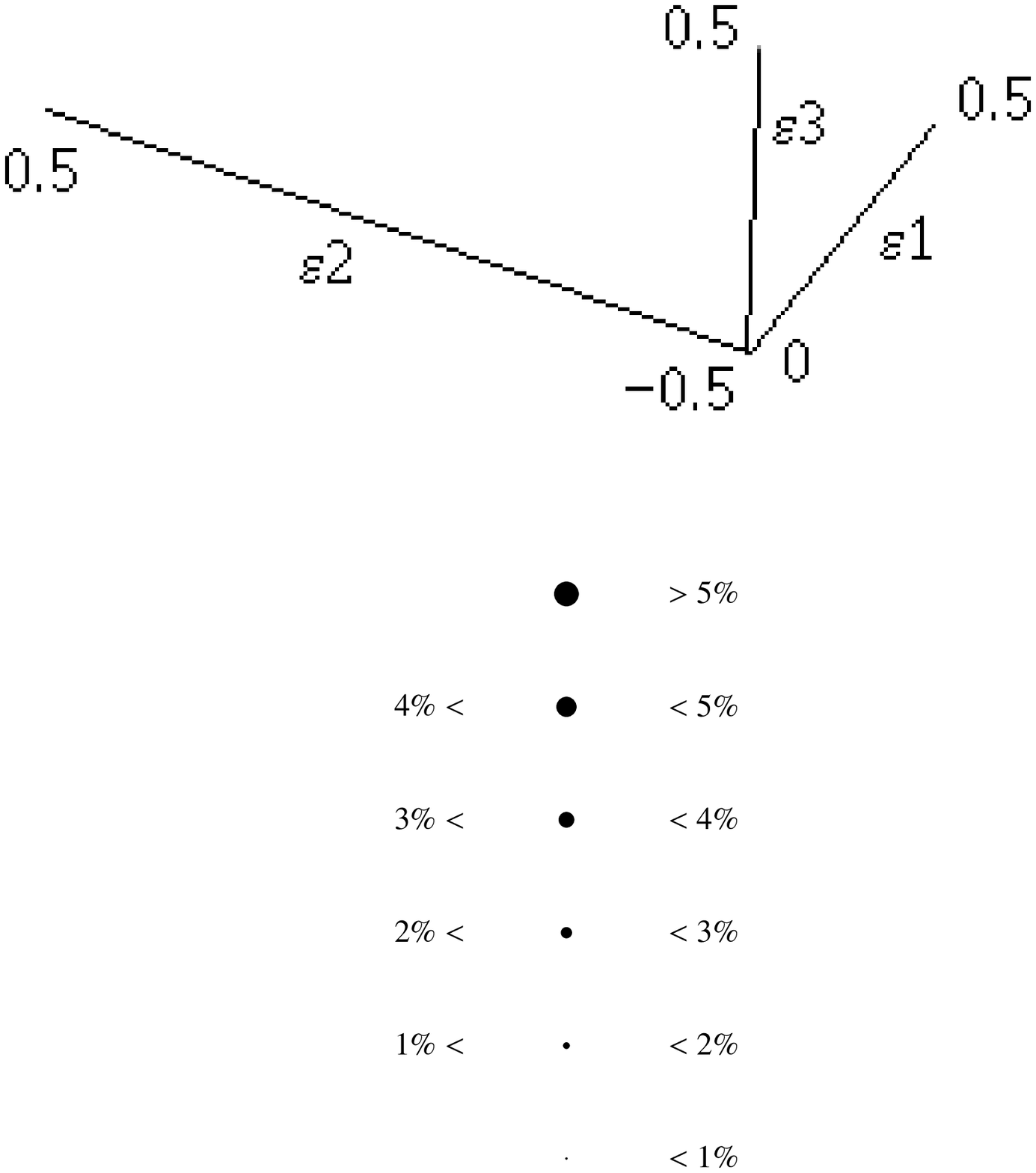}
\\
\null
\hspace{2cm}$\Delta R_{_{\rm MCE}}$
\caption{Percentage differences~(\ref{Y_X}) between scalar-to-tensor
ratios given by the GFM and those obtained from the WKB, ${\rm WKB*}$
and MCE.}
\label{compare_grap}
\end{figure}
%
\subsection{Second~slow-roll~order MCE and WKB versus~GFM}
\label{mixed}
\begin{figure}[!ht]
\includegraphics[width=0.5\textwidth]{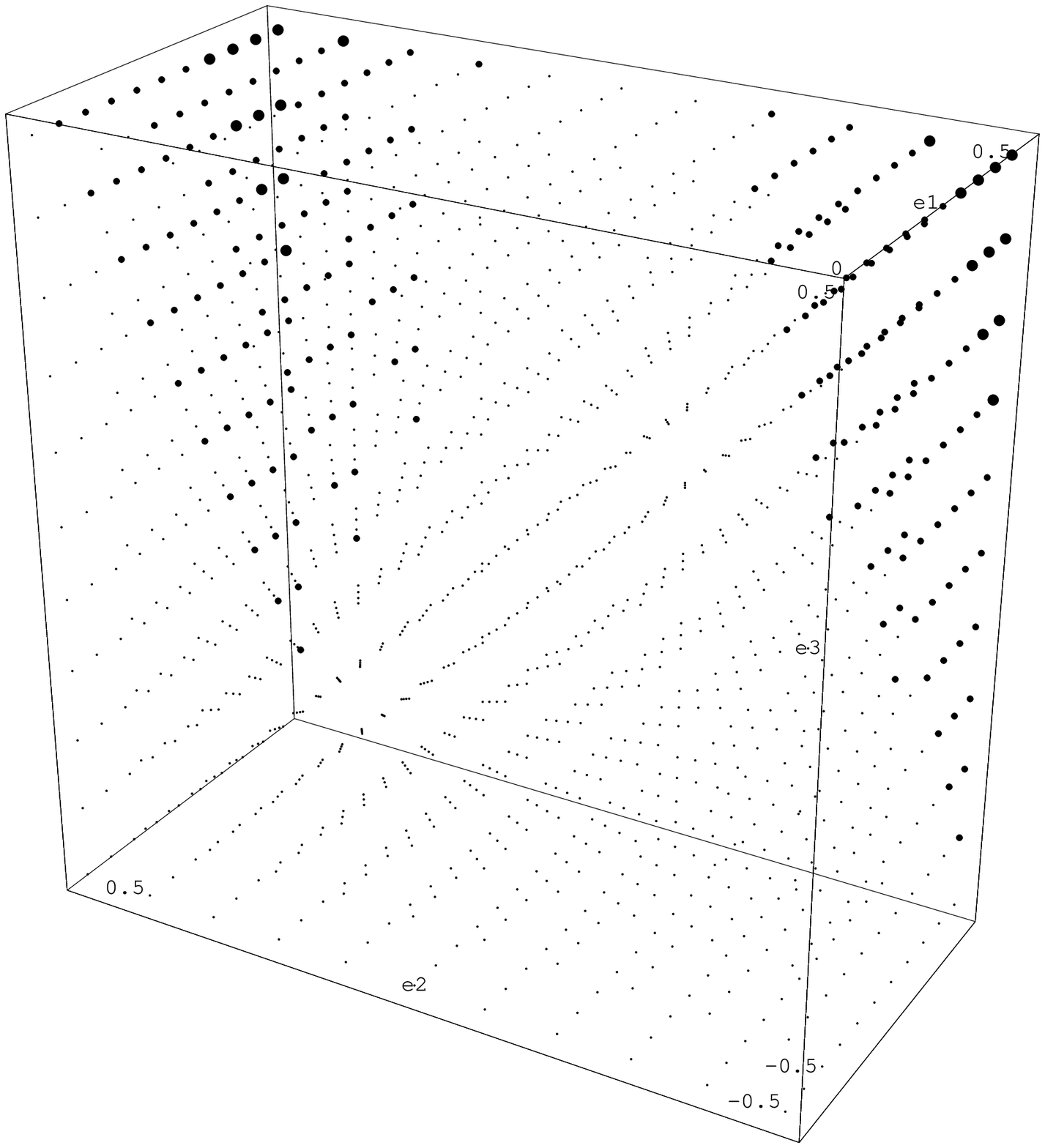}
\hspace{0.1cm}
\includegraphics[width=0.5\textwidth]{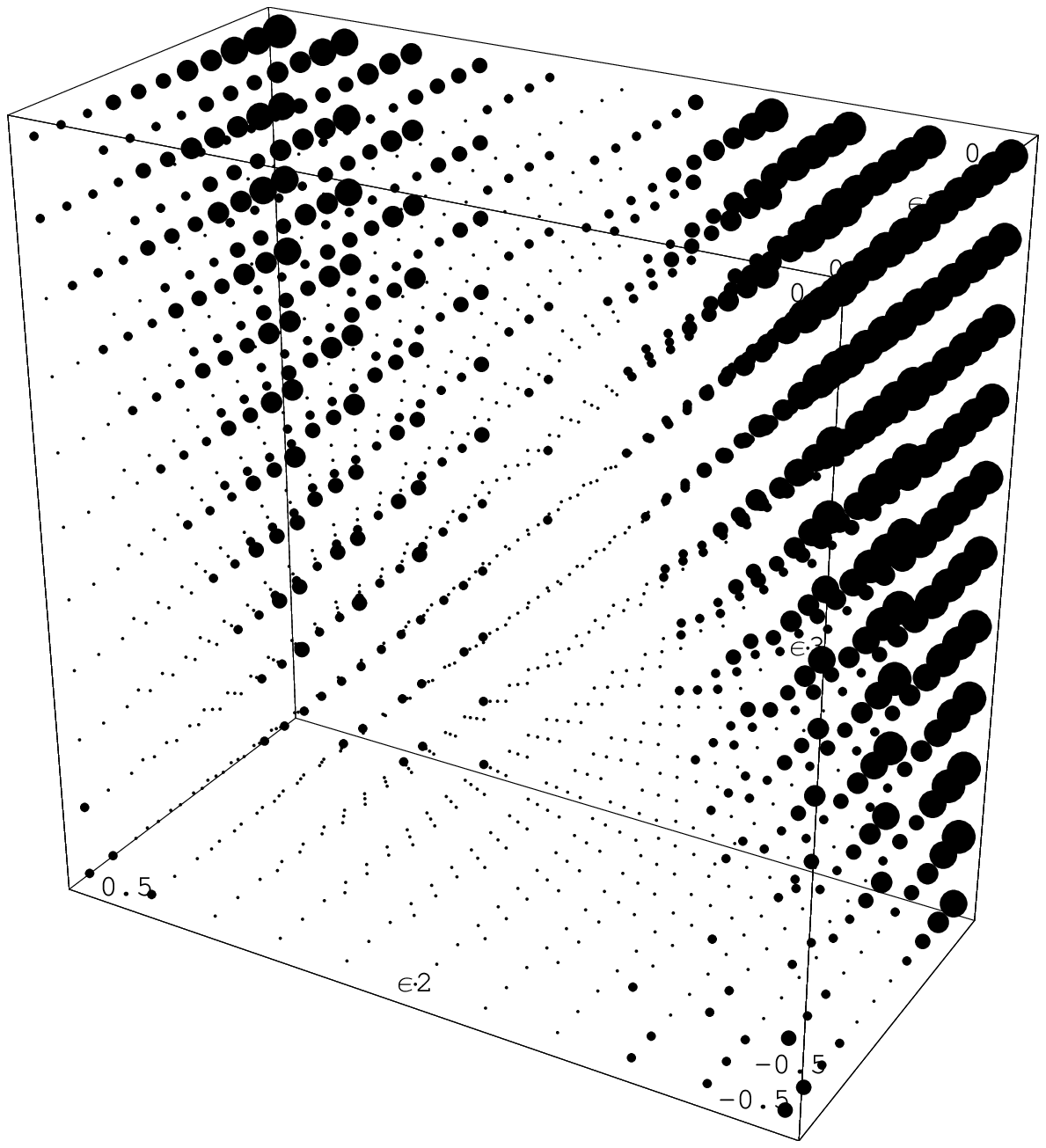}
\\
\null
\hspace{2cm} $|n_{\rm S\,\scriptscriptstyle{MCE}}-n_{\rm S\,\scriptscriptstyle{GFM}}|
\times 100$ 
\hspace{2cm} $\Delta P_{\zeta\,\scriptscriptstyle{\rm MCE}}(k_*)$
\caption{Left box: absolute difference between scalar spectral indeces $n_{\rm S}$
evaluated with the MCE and GFM (rescaled by a factor of $100$ for convenience).
Note that relevant differences of order $0.05$ (shown as $5$) occur at the
boundaries of the intervals considered for the $\epsilon_i$'s.
Right box: percentage difference~(\ref{Y_X}) for the amplitudes of scalar
perturbations $P_\zeta$ at the pivot scale $k=k_*$ between the MCE and GFM.
See Fig.~\ref{compare_grap} for the meaning of dot size.}
\label{compare_ns}
\end{figure}
\begin{figure}[h]
\includegraphics[width=0.7\textwidth]{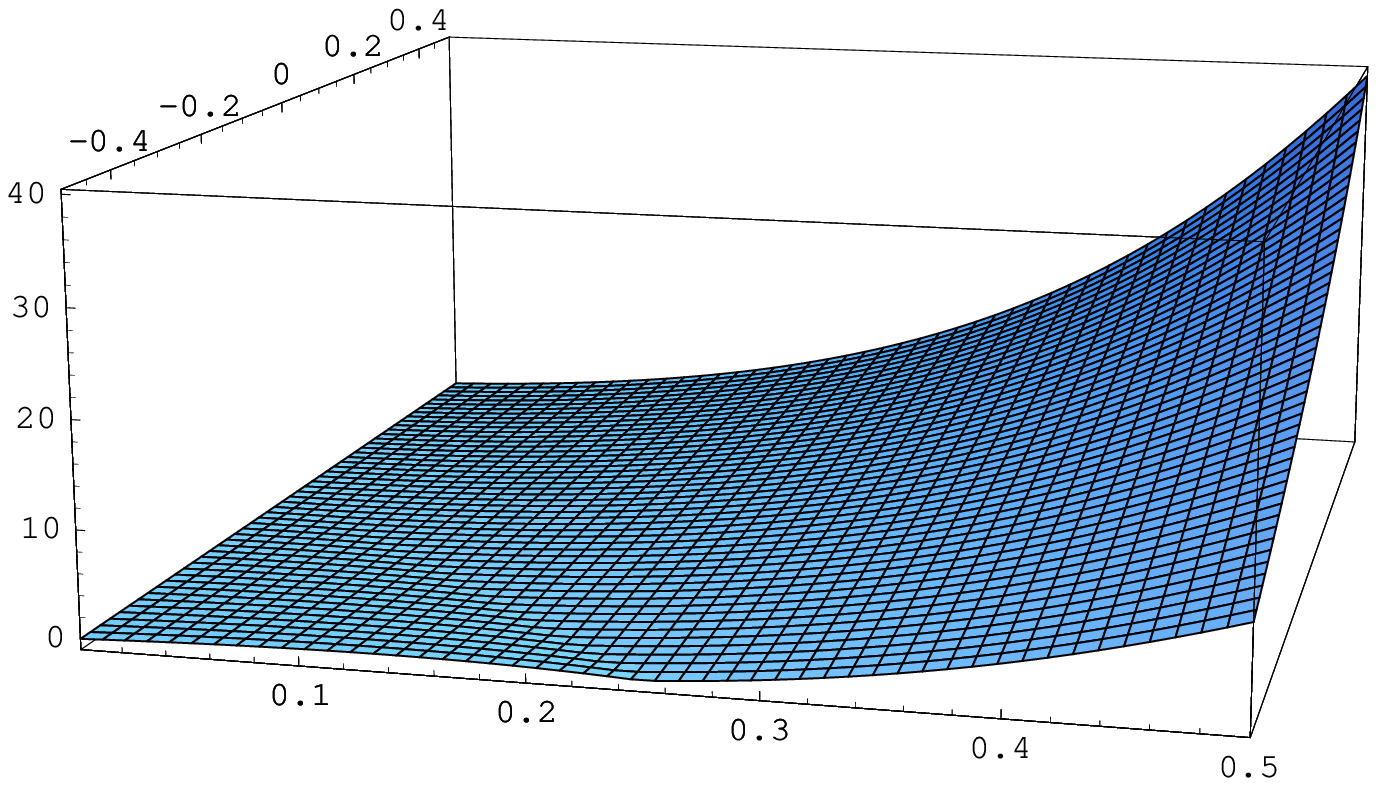}
\raisebox{3cm}{$\Delta P_{\zeta\,\scriptscriptstyle{\rm MCE}}(k_*)
|_{\epsilon_2=-2 \epsilon_1}$}
\caption{Percentage difference~(\ref{Y_X}) between the MCE and the GFM
for $P_\zeta(k_*)$ restricted on the hypersurface $\epsilon_2=-2\,\epsilon_1$,
which corresponds to a scale-invariant spectrum to first order in the slow-roll
expansion.
The graph is given for $0<\epsilon_1<0.5$ and $-0.5<\epsilon_3<0.5$.}
\label{figure_PS}
\end{figure}
We shall now compare the slow-roll results obtained from the MCE
and other methods of approximation previously
employed~\cite{WKB_lungo} with those obtained using the GFM in the
slow-roll expansion.
We use the GFM just as a reference for the purpose of comparing the
other methods among each other and of illustrating deviations from pure
slow-roll results.
For a given inflationary ``observable'' $Y$ evaluated with the method
X, we denote the percentage difference with respect to its value given
by the GFM as  
\be
\Delta Y_{_{\rm X}}
\equiv
100\,\left|\frac{Y_{_{\rm X}}-Y_{_{\rm GFM}}}{Y_{_{\rm GFM}}}\right|\%
\ ,
\label{Y_X}
\ee
where ${\rm X}={\rm WKB}$ 
stands for the first WKB and second slow-roll
orders~\cite{WKB_PLB},
${\rm X}={\rm WKB*}$ for the second WKB and second slow-roll
orders~\cite{WKB_lungo} and, of course ${\rm X}={\rm MCE}$ for the
result obtained in this paper. Note that for the case ${\rm X}={\rm 
WKB*}$ we shall set the three
undetermined parameters $b_{\rm S}=b_{\rm T}=d_{\rm S}=2$ in order
to minimize the difference with respect to the results of the GFM
(see~\ref{und_par} and Ref.~\cite{WKB_lungo}).
\par
In Fig.~\ref{compare_grap} we show, with dots of variable size,
the percentage differences~(\ref{Y_X}) for the scalar-to-tensor
ratios $R$ at the pivot scale $k=k_*$
for $0<\epsilon_1<0.5$, $|\epsilon_2|$ and $|\epsilon_3|<0.5$.
From the plots it appears that the level of accuracy of the MCE is
comparable to that of the ${\rm WKB*}$ and both are (almost everywhere)
more accurate than the WKB.
However, the MCE achieves such a precision at leading order and is thus
significantly more effective than the ${\rm WKB*}$. 
In Fig.~\ref{compare_ns} we show the difference in $n_{\rm S}$ and
the relative difference in $P_{\zeta} (k_*)$ between the MCE and GFM.
In Fig.~\ref{figure_PS} we finally plot the relative difference in
$P_{\zeta} (k_*)$ for $\epsilon_2=-2\,\epsilon_1$ (the scale 
invariant case to first order in slow-roll parameters). 
\subsection{GFM and slow-roll approximation}
Let us further compare our results with those obtained by the GFM.
As we stated before, the most general result of the present work
is given by the expressions for the power spectra~(\ref{spectra_S})
and~(\ref{spectra_T}) with the corrections~(\ref{corr_TP}).
In fact, the spectral indices and their runnings are the same as
found with the standard WKB leading order~\cite{MS,WKB_lungo}.
On the other hand, the main results of the GFM are the expressions
for power spectra, spectral indices and runnings ``$\ldots$ in the
slow-roll expansion.'' that is, for small HFF (see,~e.g.,~Eqs.~(41)
and~(43) in Ref.~\cite{gongstewart}).
Here, and in some previous work~\cite{MS,WKB1}, we instead obtain
results which hold for a general inflationary context, independently
of the ``slow-roll conditions''~\footnote{Our approximation method
does not require small HFF.}
or ``slow-roll approximation''~\footnote{We expand instead according
to the scheme given in Ref.~\cite{WKB_lungo}.}.
The slow-roll case is just a possible application of our general
expressions which can be evaluated for any model by simply specifying
the scale factor.
Our general, and non-local, expressions in fact take into account all
the ``history'' of the HFF during inflation.
For example, we note that the MCE at leading order reproduces the exact
result for power-law inflation (see Sec.~\ref{power-law}),
whereas the GFM reproduces those to the next-to-leading
order~\cite{gongstewart}.
\par
\begin{figure}[t!]
\raisebox{3cm}{$\epsilon_i$}
\includegraphics[width=0.7\textwidth]{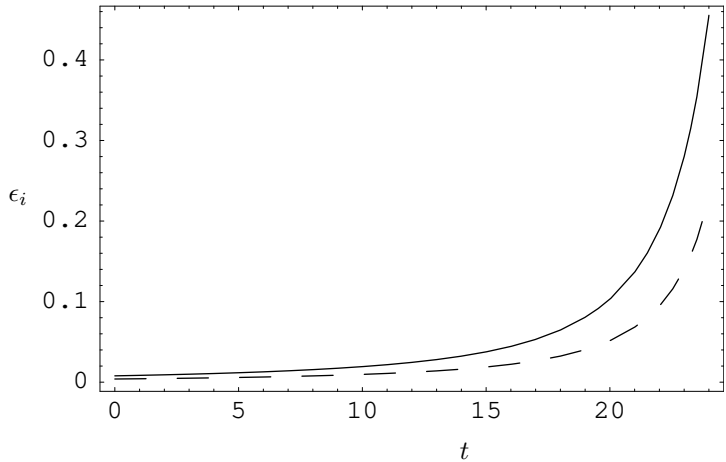}
\\
\null
\hspace{6cm}$t$
\caption{HFF in the slow-roll regime:
the dashed line is $\epsilon_1$ and the solid line is all the
$\epsilon_n$'s with $n\ge2$ (time in units of $1/m$).}
\label{eps1_epsn}
\end{figure}

Chaotic inflation~\cite{linde} is another example for which a Taylor
approximation of the HFF (such as the one required by the GFM~\cite{gongstewart}
or our Green's function perturbative expansion in~\ref{VENTURI_pert})
may lead to inaccurate results.
We consider a quadratic potential
\be
V(\phi)=\frac{1}{2}\,m^2\,\phi^2
\ ,
\label{chaos}
\ee
where $m$ is the mass of the inflaton $\phi$.
In a spatially flat Robertson-Walker background, the potential energy
dominates during the slow-rollover and the Friedman equation becomes
\be
H^2=\frac{4\,\pi}{3\,m_{\rm Pl}^2}\,\left[\dot\phi^2+m^2\,\phi^2\right]
\simeq\frac{4\,\pi\,m^2\,\phi^2}{3\,m_{\rm Pl}^2}
\ .
\label{hubble}
\ee
\begin{figure}[t!]
\raisebox{3cm}{$H$}
\includegraphics[width=0.7\textwidth]{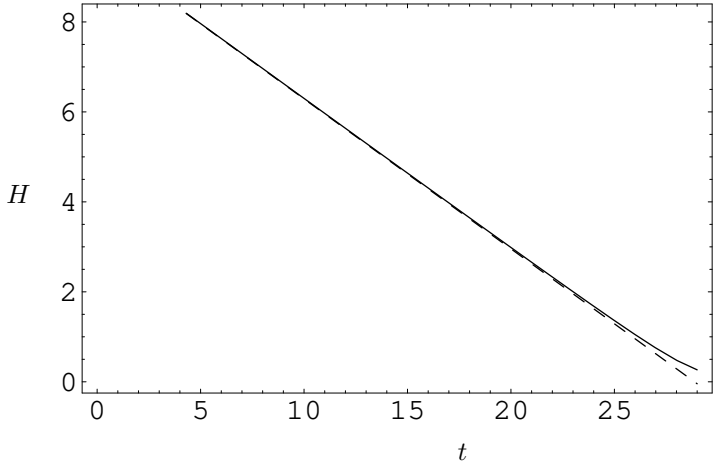}
\\
\null
\hspace{6cm}$t$
\caption{Numerical evolution of $H$ (solid line) and its
analytic approximation (dashed line).
The initial condition corresponds to $H_i\approx 8.2\,m$ (time
in units of $1/m$) when the inflaton starts in the slow-roll regime
with a value $4\,m_{\rm Pl}$.}
\label{fig:hubble}
\end{figure}
The Hubble parameter evolves as
\be
\dot H=-\frac{4\,\pi}{m_{\rm Pl}^2}\,\dot\phi^2
\ .
\label{hubbleder}
\ee
On using the equation of motion for the scalar field
\be
\ddot\phi+3\,H\,\dot\phi+m^2\,\phi=0
\ ,
\label{scalar_hom}
\ee
and neglecting the second derivative with respect to the cosmic time,
we have $3\,H\,\dot\phi\simeq-\,m^2\,\phi$.
Eq.~(\ref{hubbleder}) then yields
\be
\dot H\simeq-\frac{m^2}{3}\equiv\dot H_0
\ ,
\label{hdot}
\ee
which leads to a linearly decreasing Hubble parameter and, correspondingly,
to an evolution for the scale factor which is not exponentially linear in
time, i.e.
\numparts
\be
&&
H(t)\simeq H_0+\dot H_0\,t
\ ,
\label{evoH}
\\
&&
a(t)\simeq a(0)\,\exp\left(H_0\,t+\dot H_0\,\frac{t^2}{2}\right)
\equiv a(0)\,\exp\left[N(t)\right]
\ .
\label{evoA}
\ee
\endnumparts
In Fig.~\ref{fig:hubble} the analytic approximation~(\ref{evoH}) is
shown to be very good on comparing with the exact (numerical) evolution.
We are interested in the HFF in the slow-roll regime.
From the definitions~(\ref{HFF_def}) we then have
\numparts
\be
&&
\epsilon_1(t)=
-\frac{\dot H_0}{\left(H_0+\dot H_0\,t\right)^2}
\label{eps1_epsn_funct1}
\\
&&
\epsilon_n(t)=
-\frac{2\,\dot H_0}{\left(H_0+\dot H_0\,t\right)^2}
=2\,\epsilon_1(t)
\ ,
\quad
n\ge2
\ ,
\label{eps1_epsn_funct2}
\ee
\endnumparts
which we plot in Fig.~\ref{eps1_epsn}.
\par
\begin{figure}[t]
\raisebox{3cm}{$\delta_1$}
\includegraphics[width=0.7\textwidth]{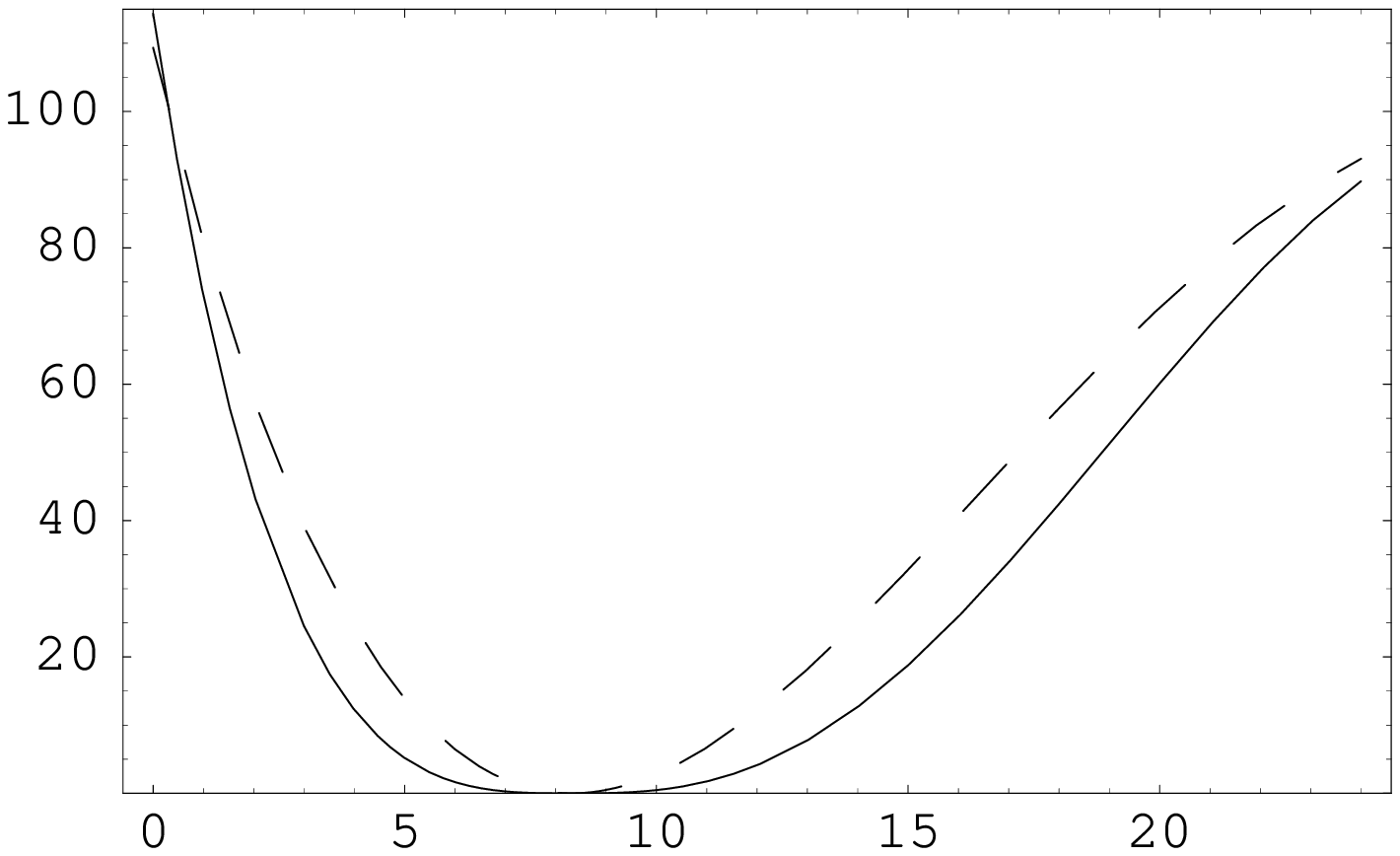}
\\
\null
\hspace{6cm}$t$
\caption{Percentage error for $\epsilon_1$ to first and second order in
$\Delta N$ (dashed and solid line respectively; time in units of $1/m$).}
\label{err1_err2}
\end{figure}

Let us take the end of inflation at $t_{\rm end}\simeq 24/m$
(so that the analytic approximation~(\ref{evoH}), (\ref{evoA}) is very
good till the end), corresponding to $N_{\rm end}=125$~e-folds.
We then consider the modes that leave the horizon 60~e-folds before
the end of inflation and take that time ($t=t_*\simeq 8/m$ or $N_*=125-60=65$)
as the starting point for a Taylor expansion to approximate the
HFF~\footnote{One can of course conceive diverse expansions, e.g.~using
other variables instead of $N$, however the conclusion would not change
as long as equivalent orders are compared.},
\be
\epsilon_i\left(\Delta N\right)\simeq
\epsilon_i+\epsilon_i\,\epsilon_{i+1}\,\Delta N
+\frac12\,\epsilon_i\,
\left(\epsilon_{i+1}^2+\epsilon_{i+1}\,\epsilon_{i+2}\right)\,
\Delta N^2
\label{epsn_expan}
\ ,
\ee
where the $\epsilon_i$'s in the r.h.s.~are all evaluated at the time
$t_*$ and $\Delta N=N-N_*$.
The percentage error with respect to the analytic
expressions~(\ref{eps1_epsn_funct1}) and~(\ref{eps1_epsn_funct2}) ,
\be
\delta_i\equiv100\,\left|\frac{\epsilon_i(t)
-\epsilon_i\left(\Delta N\right)}{\epsilon_i(t)}
\right|\%
\label{err_epsn_expan}
\ ,
\ee
with $\Delta N=N(t)-N_*$ is then plotted in Fig.~\ref{err1_err2}
for $i=1$.
It obviously becomes large rather quickly away from $t=t_*$ and
we can immediately conclude that, had we used a Taylor expansion to approximate
the HFF over the whole range of chaotic inflation ($-65<\Delta N<60$), we would
have obtained large errors from the regions both near the beginning and the end of
inflation.
\par
In general, we expect that the Taylor expansion of the HFF will lead to 
a poor determination of the numerical coefficients (depending on $C$ in
Eqs.~(\ref{PS_SlowRoll_0_2order}) and (\ref{PT_SlowRoll_0_2order})) in 
the second slow-roll order terms for those models in which the HFF 
vary significantly in time. 
In fact, we can calculate the integral of Eq.~(21) in Ref.~\cite{gongstewart}
with $\epsilon_1$ instead of the complete $g\left(\ln x\right)$ and $y_0(u)$
instead of $y(u)$.
The percentage difference between this integral calculated using the Taylor
expansion~(\ref{epsn_expan}) with respect to the same integral calculated with
$\epsilon_1$ in Eqs.~(\ref{eps1_epsn_funct1}) and~(\ref{eps1_epsn_funct2}) is
$92\,\%$ and $88\,\%$ to first and second orders in $\Delta N$, respectively.
The MCE method does not require such an expansion and does therefore not suffer
from this restriction~\footnote{We again refer the reader to Section~III of
Ref.~\cite{WKB_lungo} for more details.}.
\par
\begin{figure}[t!]
\raisebox{3cm}{${\rm Log}(P_Q)$}
\includegraphics[width=0.7\textwidth]{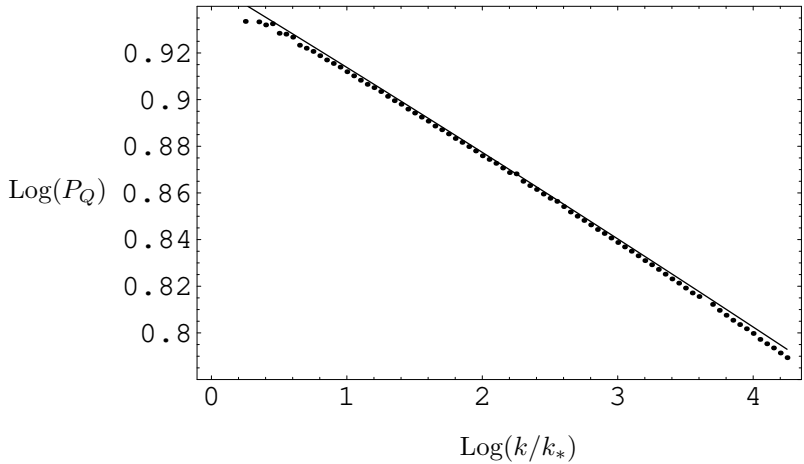}
\\
\null
\hspace{6cm}${\rm Log}(k/k_*)$
\caption{Spectrum of the Mukhanov variable $Q$ ($P_Q=k^3\,|Q_k|^2/(2\,\pi^2)$)
evaluated at the end of inflation for the chaotic model of Eq.~(\ref{chaos}).
The dots represent the numerical values and the solid line the analytic
fit based on Eqs.~(\ref{n_2'order_S}) and (\ref{alpha_2'order_S}) 
obtained by the second slow-roll order MCE approximation.
The first order and GFM analytic results are not shown 
since they are almost indistinguishable from the MCE result plotted.
$k_*$ crosses the Hubble radius at $\phi_*\simeq 3\,m_{\rm Pl}$
(the lines are normalized at $10^{2.25}\,k_*$).}
\label{fig:spe}
\end{figure}
We have also compared our analytic MCE results in Eqs.~(\ref{n_2'order_S})
and (\ref{alpha_2'order_S}) with the numerical evaluation of the spectrum
for scalar perturbations. 
In particular, we have evolved in cosmic time the modulus of the Mukhanov
variable $Q$ (recalling that $P_\zeta=k^3\,|Q_k|^2/(2\,\pi^2)$), 
which satisfies the associated non-linear Pinney equation~\cite{BFV}.
The initial conditions for the numerical evolution are fixed for 
wave-lengths well within the Hubble radius and correspond to the adiabatic 
vacuum, i.e.~$|Q_k(t_i)|=1/(a(t_i)\,\sqrt{k})$ and
$|\dot Q_k|(t_i) = - H(t_i)\,|Q_k(t_i)|$~\cite{FMVV_1}. 
The agreement of the analytic MCE approximation with the numerical results
is good, as shown in Fig.~\ref{fig:spe}. 
\subsection{Beyond slow-roll}
\begin{figure}[t!]
\raisebox{3cm}{$\epsilon_i$}
\includegraphics[width=0.7\textwidth]{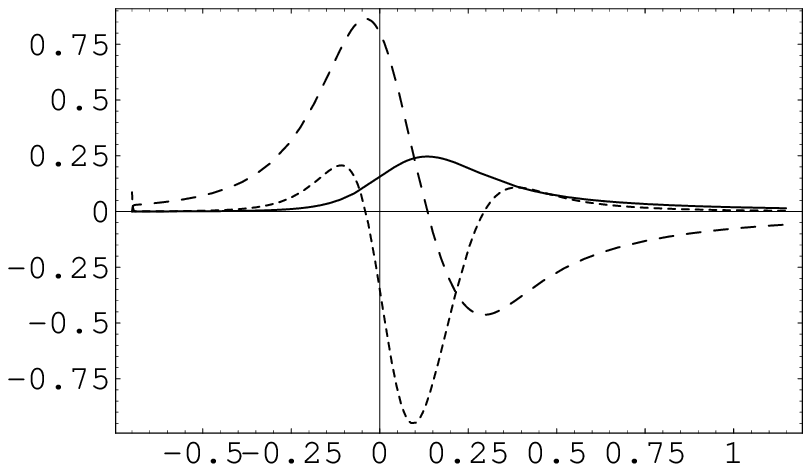}
\\
\null
\hspace{6cm} $\phi/m_{\rm Pl}$
\caption{Evolution of $\epsilon_i$ with the value of the inflaton
$\phi$ (in units of $m_{\rm Pl}$) in the arctan model of
Eq.~(\ref{arctan}):
$\epsilon_1$ (solid line), $\epsilon_2$ (long-dashed line) and
$\dot\epsilon_2/H=\epsilon_2\,\epsilon_3$ (short-dashed line).} 
\label{arctan_figure1}
\end{figure}

\begin{figure}[t!]
\raisebox{3cm}{${\rm Log}(P_Q)$}
\includegraphics[width=0.7\textwidth]{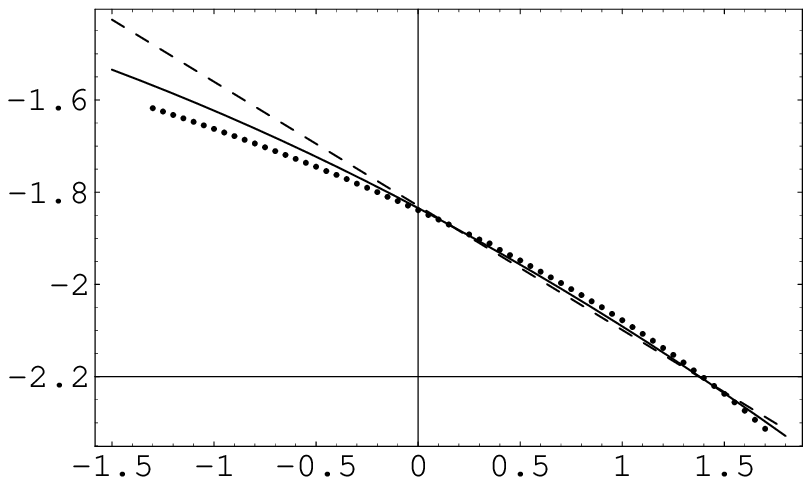}
\\
\null
\hspace{6cm} ${\rm Log} (k/k_*)$
\caption{Spectrum of the Mukhanov variable $Q$ ($P_Q=k^3\,|Q_k|^2/(2\,\pi^2)$)
for the arctan model of Eq.~(\ref{arctan}).
The dots represent the numerical values, the solid line the analytic
results from Eqs.~(\ref{n_2'order_S}) and (\ref{alpha_2'order_S})
obtained by the second slow-roll order MCE approximation and
the dashed line those given by the first order slow-roll approximation.
$k_*$ crosses the Hubble radius at $\phi_*\simeq -0.3\,m_{\rm Pl}$
(the lines are normalized at $10^{3/20} k_*$) and the spectrum is
evaluated at $\simeq 55$~e-folds afterwards.} 
\label{arctan_figure2}
\end{figure}

The slow-roll approximation is quite accurate for a wide class of 
potentials.
However, violations of the slow-roll approximation may occur 
during inflation, leading to interesting observational effects in the
power spectra of cosmological perturbations.
An archetypical model to study such violations is given by the potential
\be
V(\phi)=V_0\,\left[1
-\frac{2}{\pi}\,\arctan\left(N\,\frac{\phi}{m_{\rm Pl}}\right)\right]
\ ,
\label{arctan}
\ee
introduced with $N=5$ in Ref.~\cite{WMS}.
In Fig.~\ref{arctan_figure2} we show that the MCE result expanded to
second slow-roll order provides a very good fit for the power spectrum
of scalar perturbations even in situations where the slow-roll parameters
are not very small (see Fig.~\ref{arctan_figure1}).
This example also shows that second slow-roll order results are much
better than first order ones (analogous results were also obtained with
the GFM in Ref.~\cite{LLMS}). 
\section{Conclusions}
\label{sC}
We have presented the application of the method of comparison equations to 
cosmological perturbations during inflation.
By construction (i.e.~by choosing a suitable comparison function), this 
approach leads to the exact solutions for inflationary models with constant
horizon flow functions $\epsilon_i$'s (e.g.~power-law inflation). 
\par
The main result is that, on using this approach to leading order, we were
able to obtain the general expressions~(\ref{spectra_S})-(\ref{corr_TP})
for the inflationary power spectra which are more accurate than
those that any other method in the literature can produce at the
corresponding order.
In fact, the MCE leads to the correct asymptotic behaviours (in contrast with
the standard slow-roll approximation~\cite{SL}) and solves the difficulties in
finding the amplitudes which were encountered with the WKB method~\cite{MS}. 
\par
Starting from the general results~(\ref{spectra_S})-(\ref{corr_TP}),
we have also computed the full analytic expressions for the inflationary power
spectra to second slow-roll order in Eqs.~(\ref{PS_SlowRoll_0_2order})
and~(\ref{PT_SlowRoll_0_2order}) and found that the dependence on the
horizon flow functions $\epsilon_i$'s is in agreement with that obtained
by different schemes of approximation, such as the GFM~\cite{gongstewart} and
the WKB approximation~\cite{WKB_PLB,WKB_lungo}.
Moreover, the results obtained with the MCE do not contain undetermined
coefficients, in contrast with second slow-roll order results obtained
with the WKB$*$~\cite{WKB_lungo}. 
\par
Let us conclude by remarking that, just like the WKB approach, the MCE does
not require any particular constraints on the functions $\epsilon_i$'s
and therefore has a wider range of applicability than any method which
assumes them to be small.
As an example, we have discussed in some detail the accuracy of the MCE for
the massive chaotic and arctan inflationary models.
We have shown that the MCE leads to accurate predictions even for a model
which violates the slow-roll approximation during inflation. 
\ack
We would like to thank S.~Leach for discussions, A.~O.~Barvinsky and G.~P.~Vacca
for discussions and comments on the manuscript.
\appendix
\section{Impact of undetermined parameters in the WKB* approximation}
\label{und_par}
\begin{figure}[hb!]
\includegraphics[width=.7\textwidth]{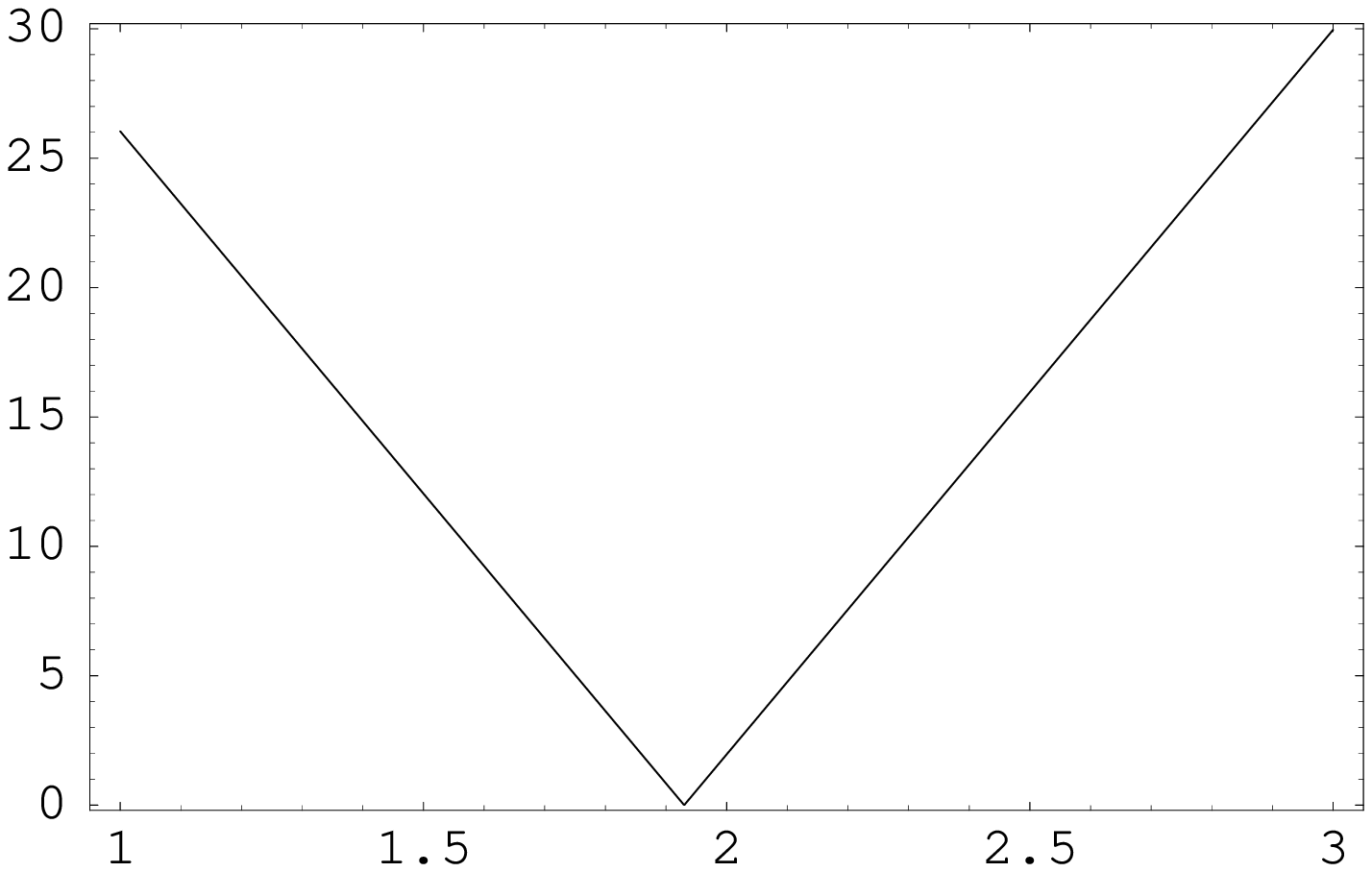}
\\
\null
\hspace{5cm}$b_{\rm T}$
\caption{Percentage difference between the numerical coefficients of
$\epsilon_1\,\epsilon_2$ in $P_h$ from ${\rm WKB*}$ and ${\rm GFM}$ as
a function of $b_{\rm T}$.
The difference vanishes for $b_{\rm T}\simeq 1.93$ ($\simeq 2$, the value
used in the main text).}
\label{param_plot}
\end{figure}

To clarify the impact of the undetermined parameters
$b_{\rm S}$, $b_{\rm T}$, and $d_{\rm S}=2$ on the
results to second order in the HFF, the percentage differences
between the numerical coefficients of some second order terms in the
power spectra are displayed in Table~\ref{perc_diff_tab_PS} and of some
of those in the scalar-to-tensor ratio in Table~\ref{perc_diff_tab_R}.
All the entries are for $b_{\rm S}=b_{\rm T}=d_{\rm S}=2$, as is used in
the main text, since other choices would give larger differences, as we
explicitly show in Fig.~\ref{param_plot} for the numerical coefficient
in front of $\epsilon_1\,\epsilon_2$ in $P_h$ (since the behaviour is
the same for all coefficients, we do not display other plots).
\begin{table}[!ht]
\begin{tabular}{|c|c|c|c|c|}
\hline
&$\epsilon_1^2$; $P_\zeta$, $P_h$
&$\epsilon_2^2$; $P_\zeta$
&$\epsilon_1\,\epsilon_2$; $P_\zeta (b_{\rm S}=2)$
&$\epsilon_1\,\epsilon_2$; $P_h (b_{\rm T}=2)$
\\
\hline
${\rm WKB}$
&$2.3\%$
&$3.0\%$
&$353.3\%$
&$37.2\%$
\\
\hline
${\rm WKB*}$
&$0.03\%$
&$0.06\%$
&$29.1\%$
&$2.0\%$
\\
\hline
\end{tabular}
\caption{Percentage differences between the numerical coefficients
of some second order terms in the power spectra given by ${\rm WKB}$
and ${\rm WKB*}$ with respect to those obtained by the GFM.}
\label{perc_diff_tab_PS}
\end{table}
\begin{table}[!ht]
\begin{tabular}{|c|c|c|c|}
\hline
&$\epsilon_2\,\epsilon_3$; $(d_{\rm S}=2)$
&$\epsilon_1\,\epsilon_2$; $(b_{\rm S}-b_{\rm T}=0)$
&$\epsilon_2^2$
\\
\hline
${\rm WKB}$
&$56.7\%$
&$1.5\%$
&$117.3\%$
\\
\hline
${\rm WKB*}$
&$1.3\%$
&$0.03\%$
&$1.5\%$
\\
\hline
\end{tabular}
\caption{Percentage differences between the numerical coefficients of
some second order terms in the scalar-to-tensor ratio given by ${\rm WKB}$
and ${\rm WKB*}$ with respect to those obtained from GFM.
Note that the numerical coefficient of $\epsilon_2\,\epsilon_3$ in $R$
is the same as that in $P_\zeta$.
Further, the coefficient of $\epsilon_1\,\epsilon_2$ does not depend
on any undetermined parameters if $b_{\rm S}=b_{\rm T}$.}
\label{perc_diff_tab_R}
\end{table}
\section{Perturbative approximations}
\label{VENTURI_pert}
In this Section we present a possible method of obtaining
the complete result to second order in the HFF.
We start by noting that any given function $f=f(x)=f(\epsilon,x)$ can
be written as the power series
\be
\!\!\!\!\!\!\!
f(\epsilon,x)
=
\sum_{n=0}^\infty\,\frac{x^n}{n!}\,
\left(-\sum_i\,\frac{\hat\epsilon_i\,\hat\epsilon_{i+1}}{1-\hat\epsilon_1}\,
\frac{\partial}{\partial\hat\epsilon_i}\right)^n
f(\hat\epsilon_i,x)
\equiv
\sum_{n=0}^\infty\,\frac{x^n}{n!}\,f_n(x)
\ ,
\label{f_def}
\ee
where $\hat\epsilon_i=\epsilon_i(x_*)$ are the HFF evaluated at the
horizon crossing (i.e.~at $x=x_*$) and 
$f(\hat\epsilon_i,x)=f(\epsilon_i=\hat\epsilon_i,x)$.
On using Eqs.~(\ref{f_def}) to express both $\omega^2$ and
$\chi$ in Eq.~(\ref{exact_EQ}), we obtain
\be
\!\!\!\!\!\!\!\!\!\!\!\!\!\!\!\!\!\!\!\!
\chi_0(x)\,\sum_{n=1}^\infty\,\frac{x^n}{n!}\,\omega^2_n(x)
+\left[\frac{{\d}^2}{{\d}x^2}
+\sum_{m=0}^\infty\,\frac{x^m}{m!}\,\omega^2_m(x)
\right]\,
\sum_{n=1}^\infty\,\frac{x^n}{n!}\,\chi_n(x)
=0
\ ,
\label{exact_EQ_ter}
\ee
where we have used
\be
\left[\frac{{\d}^2}{{\d}x^2}+\omega^2_0(x)\right]
\,\chi_0(x)=0
\ ,
\label{0th_EQ}
\ee
with
$\omega^2_0(x)=\omega^2(\epsilon_i=\hat{\epsilon}_i;x)$
and $\chi_0(x)$ is a linear combination (i.e.~the MCE leading order solution)
of Eq.~(\ref{exact_SOL_bis}) for $\Theta^2(x)=\omega^2_0(x)$ and
$\sigma\to x$ in the argument of the Bessel functions.
One may solve Eq.~(\ref{exact_EQ_ter}) for higher orders by introducing
\be
\frac{x^n}{n!}\,\chi_n(x)\equiv g_n(x)\,\chi_0(x)
\ ,
\label{def_to_solve}
\ee
and using the standard formulae for second order differential equations with a
source term, so that
\be
\!\!\!\!\!\!\!\!\!\!\!\!\!\!\!\!\!\!\!\!
\!\!\!\!\!\!\!\!\!\!\!\!\!\!\!\!\!\!\!\!
g_n(x)=
-\int_{\infty}^x\,\frac{\d y}{\chi^2_0(y)}\,
\int_{\infty}^{y}\,\d z\,\chi^2_0(z)
\left[
\frac{z^n}{n!}\,\omega^2_n(z)
+\sum_{m=1}^{n-1}\,
g_m(z)\,\frac{z^{n-m}}{n-m}\,\omega^2_{n-m}(z)
\right]
\ .
\label{g_n_GEN}
\ee
The problem has therefore been reduced to computing $2\,n$ integrals.
\par
The above formalism can be modified in order to provide an
alternative way for calculating the power spectra,
spectral indices and runnings to second order in the HFF.
We start from Eq.~(\ref{osci}) and (\ref{freq}) and the exact
relations
%
\be
\frac{z''_{\rm T}}{z_{\rm T}}
=
a^{2}\,H^{2}\,\left(2-\epsilon_1\right)
\nonumber
\\
\\
\frac{z''_{\rm S}}{z_{\rm S}}
=
a^{2}\,H^{2}\,\left(2-\epsilon_1+\frac{3}{2}\,\epsilon_2
-\frac{1}{2}\,\epsilon_{1}\,\epsilon_2
+\frac{1}{4}\,\epsilon_2^2
+\frac{1}{2}\,\epsilon_2\,\epsilon_3\right)
\ .
\nonumber
\ee
%
With a redefinitions of the wave-function and variable
\be
\psi={\rm e}^{-x/2}\,\mu
\ ,
\quad\quad
\tilde{x}=\ln\left(-k\,\eta\right)
\ee
we obtain
\be
\left[\frac{{\d}^2}{{\d}\tilde{x}^2}
+\tilde\omega^2(\tilde x)\right]\,\psi(\tilde x)
=0
\ ,
\ee
where
\be
\tilde\omega^2_{\rm T}=
{\rm e}^{2\,\tilde{x}}-\frac14
-\eta^2a^2H^2\left(2-\epsilon_1\right)
\nonumber
\\
\\
\!\!\!\!\!\!\!\!\!\!\!\!\!\!\!\!\!\!\!\!\!
\tilde\omega^2_{\rm S}=
{\rm e}^{2\,\tilde{x}}-\frac14
-\eta^2a^2H^2\left(2-\epsilon_1+\frac{3}{2}\,\epsilon_2-\frac{1}{2}\,\epsilon_{1}\,\epsilon_2
+\frac{1}{4}\,\epsilon_2^2+\frac{1}{2}\,\epsilon_2\,\epsilon_3\right)
\ .
\nonumber
\ee
%
On now using the approximate relation
\be
\eta\simeq-\frac{1}{a\,H}\,
\left(1+\epsilon_1+\epsilon_1^2+\epsilon_1\,\epsilon_2\right)
\ ,
\ee
the frequencies become
%
\be
\tilde\omega^2_{\rm T}\simeq
{\rm e}^{2\,\tilde{x}}-
\left(\frac94+3\,\epsilon_1+4\,\epsilon_1^2+4\,\epsilon_1\epsilon_2\right)
\nonumber
\\
\\
\!\!\!\!\!\!\!\!\!\!\!\!\!\!\!\!\!\!\!\!
\tilde\omega^2_{\rm S}\simeq
{\rm e}^{2\,\tilde{x}}-
\left(\frac94+3\,\epsilon_1+\frac32\,\epsilon_2+4\,\epsilon_1^2
+\frac{13}{2}\,\epsilon_1\epsilon_2+\frac14\,\epsilon_2^2
+\frac12\,\epsilon_2\epsilon_3\right)
\ ,
\nonumber
\ee
%
where we only keep the second order terms in the (still
$\tilde{x}$-dependent) HFF.
We can now expand linearly each HFF around the horizon
crossing to obtain
%
\be
\!\!\!\!\!\!\!\!\!\!\!\!\!\!\!\!\!\!\!\!\!\!\!\!\!\!\!\!\!\!\!\!\!\!\!\!\!\!\!\!
\frac{{\d}^2\,\psi_{\rm T}}{{\d}\,\tilde{x}^2}
\!+\!\left[{\rm e}^{2\,\tilde{x}}\!-\!\left(\frac94
+3\,\hat{\epsilon}_1+4\,\hat{\epsilon}_1^2+4\,\hat{\epsilon}_1\hat{\epsilon}_2\right)\right]
\!
\psi_{\rm T}=-3\,\hat{\epsilon}_1\,\hat{\epsilon}_2\,\tilde{x}\,\psi_{\rm T}
\nonumber
\\
\\
\!\!\!\!\!\!\!\!\!\!\!\!\!\!\!\!\!\!\!\!\!\!\!\!\!\!\!\!\!\!\!\!\!\!\!\!\!\!\!\!
\frac{{\d}^2\,\psi_{\rm S}}{{\d}\,\tilde{x}^2}
\!+\!\left[{\rm e}^{2\,\tilde{x}}\!-\!\left(\frac94
+3\,\hat{\epsilon}_1+\frac32\,\hat{\epsilon}_2+4\,\hat{\epsilon}_1^2
+\frac{13}{2}\,\hat{\epsilon}_1\hat{\epsilon}_2+\frac14\,\hat{\epsilon}_2^2
+\frac12\,\hat{\epsilon}_2\hat{\epsilon}_3\right)\right]\!
\psi_{\rm S}
\nonumber
\\
=
-3\left(\hat{\epsilon}_1\,\hat{\epsilon}_2
+\frac12\,\hat{\epsilon}_2\,\hat{\epsilon}_3\right)\tilde{x}\,\psi_{\rm S}
\ ,
\nonumber
\ee
%
which, beside the use of $\tilde x$ instead of $x$ and the appearance of
$\tilde\omega$ instead of $\omega$, is equivalent to the expansion in
Eq.~(\ref{f_def}) for $f=\tilde\omega^2$ and can be solved by means of a
standard perturbative approach (such as in Ref.~\cite{gongstewart}).
The method is close to GFM and gives the same results of Eq.~(41) in
Ref.~\cite{gongstewart} and Eq.~(29) in Ref.~\cite{LLMS} for the scalar and
tensor power spectra, respectively.
\par
Let us finally note that, for the GFM, the scalar spectral index and its running
are calculated by taking the derivative of Eq.~(41) in Ref.~\cite{gongstewart},
i.e.~the power spectra evaluated at $k=a\,H$, in terms of $\ln k$, while in our MCE
(but also in the WKB approximation~\cite{WKB_PLB,WKB_lungo}) the 
dependences
in $\ln\left(k/k_*\right)$ are obtained explicitly (see, for example,
Eqs.~(\ref{PS_SlowRoll_0_2order}) and (\ref{PT_SlowRoll_0_2order})).
\section*{References}

\end{document}